\documentclass[
aps, 
pra,
superscriptaddress,
showpacs,
showkeys,
notitlepage,
twocolumn, 
numbers,
longbibliography,
nofootinbib]
{revtex4-1}

\usepackage[T1]{fontenc} % alphabets to prepare
\usepackage[utf8]{inputenc} % input encoding
\usepackage[english]{babel} % language(s)
\usepackage{graphicx, xcolor}
\usepackage{amsfonts, amssymb, amsmath, mathrsfs, calc, array, mathdesign, mathtools}
\usepackage[separate-uncertainty=true]{siunitx}
\usepackage{hyperref}
\usepackage{abraces}
\renewcommand{\arraystretch}{0.45}

\usepackage{tabstackengine}
\stackMath
\newcolumntype{C}[1]{>{\centering\arraybackslash$}m{#1}<{$}}
   
\def\bra#1{\mathinner{\langle{#1}|}}
\def\ket#1{\mathinner{|{#1}\rangle}}

\def\ontop#1#2{\setbox0\hbox{#2}\copy0\llap{\raise\ht0\hbox{#1}}}
\definecolor{darkblue}{rgb}{0,0,0.93} 
\definecolor{darkred}{rgb}{0.8,0,0} 
\hypersetup{
colorlinks=true, 
linkcolor=darkblue, 
citecolor=darkred, 
filecolor=darkblue, 
urlcolor=darkblue
}

\newenvironment{poliabstract}[1]
   {\renewcommand{\abstractname}{#1}\begin{abstract}}
   {\end{abstract}}

\begin{document}

\title{Discrete-time quantum walks as fermions of lattice gauge theory}

\author{Pablo Arnault}
\email{pablo.arnault@ific.uv.es}
\affiliation{Departamento de F{\'i}sica Te{\'o}rica and IFIC, Universidad de Valencia and CSIC, Dr.\ Moliner 50, 46100 Burjassot, Spain}

\author{Armando P{\'e}rez}
%\email{armando.perez@uv.es}
\affiliation{Departamento de F{\'i}sica Te{\'o}rica and IFIC, Universidad de Valencia and CSIC, Dr.\ Moliner 50, 46100 Burjassot, Spain}

\author{Pablo Arrighi}
%\email{pablo.arrighi@lif.univ-mrs.fr}
\affiliation{Aix-Marseille Univ., Universit{\'e} de Toulon, CNRS, LIS, Marseille and IXXI, Lyon, France}

\author{Terry Farrelly}
%\email{farreltc@tcd.ie}
\affiliation{Institut f\"ur Theoretische Physik, Leibniz Universit\"at Hannover, Appelstr.\ 2, 30167 Hannover, Germany}

\begin{poliabstract}{abstract}
It is shown that discrete-time quantum walks can be used to digitize, i.e., to time discretize fermionic models of continuous-time lattice gauge theory.
The resulting discrete-time dynamics is thus not only manifestly unitary, but also ultralocal, i.e.\ the particle's speed is upper bounded, as in standard relativistic quantum field theories.
The lattice chiral symmetry of staggered fermions, which corresponds to a translational invariance, is lost after the requirement of ultralocality of the evolution; this fact is an instance of Meyer's 1996 no-go lemma stating that no non-trivial one-dimensional scalar quantum cellular automaton can be translationally invariant \cite{Meyer96a}.
All results are presented in a single-particle framework and for a (1+1)-dimensional spacetime.
\end{poliabstract}



\maketitle

\paragraph*{{Introduction.}}

Lattice gauge theories (LGTs) are a framework to define and study non-perturbative quantum field theory (QFT) \cite{book_Rothe}. 
The associated computations being often very demanding, they are usually evaluated via Monte-Carlo sampling, but this technique is limited for the computationally- most-demanding questions, such as real-time dynamics, and it suffers from the sign problem \cite{Gattringer2016}.
Tensor-network techniques can improve \cite{Sandvik2007, Schuch2008} or circumvent and outperform \cite{Buyens2014b,Banuls2014,Banuls2017} Monte-Carlo sampling, but one still expects at the very least a substantial speed-up from quantum computers \cite{Markov2008}.
Now, although a fully-fledged quantum computer is not yet to be built, various types of quantum simulators have already provided proofs of principle, from chemistry, to condensed matter, to high-energy physics.
Regarding the latter, a small-size digital quantum simulator based on trapped ions was, indeed, recently built, which successfully reproduced pair creation in the Schwinger model \cite{Martinez16}.
Also, several proposals of analog \cite{Wiese2013, Zohar2015, ZHJOBH18} and digital \cite{Zohar_2017} quantum simulations of LGTs with cold atoms in optical lattices have been made in the last years.
The model implemented in Martinez et al.'s experiment \cite{Martinez16} is a unitary digitization, i.e., time discretization, via a series of quantum gates, of a continuous-time formulation of (1+1)D lattice quantum electrodynamics.
This continuous-time model can actually be seen as a continuous-time quantum walk (CTQW) having multiparticle gauge interactions, with fermionic fields on the sites of the lattice and bosonic fields on the links.
Quantum walks, be they in continuous or discrete time, are models of quantum transport on graphs, e.g.\ spatial lattices, which are useful both to design quantum algorithms and for quantum simulation.
They have actually been suggested for universal quantum computation \cite{CGW13}.

Discrete-time quantum walks (DTQWs) are defined in discrete time. The state of the walker at time $j+1 \in\mathbb{N}^{\ast}$, and position $p \in \mathbb{Z}$, is determined solely by the state of the walker at time $j$ \emph{within a certain bounded spatial neighborhood around $p$}.
This defines ultralocality for an evolution operator evolving states in discrete time.
Multiparticle DTQWs are known as quantum cellular automata (QCA) \cite{Vogts09}.
%
%We will use the wording `ultralocal evolution' for `QCA' or `DTQW'.
%
Ultralocality of the evolution, not only spares resources, but actually preserves a fundamental property of standard relativistic QFTs in continuous spacetime: the existence of a maximum speed, i.e., an upper bound on the propagation speed.
This feature is at the heart of a structure theorem which proves that any QCA can be built out of a small subfamily of QCA, in a way which preserves spacetime neighborhoods \cite{Arrighi2012}.
QCA thus seem natural candidates to discretize relativistic QFTs.
Several results have already been obtained, (i) in the single-particle free case \cite{Arrighi_higher_dim_2014, Farrelly2014a, Farrelly2014b}, with couplings to electromagnetic \cite{AD16}, non-Abelian \cite{ADMDB16} and relativistic gravitational gauge fields \cite{AD17,AF17}, and (ii) in the multiparticle free case \cite{Vogts09, Farrelly15, DAriano2016b}.
Some results exist in the multiparticle interacting case \cite{Yepez2016}.
Also, action principles for QCA and their general-relativistic covariance have been studied \cite{Arrighi2014,Yepez2016,ADMF16,D18}.
That being said, there has been little work, even in the single-particle case, on the relationship between the well-known discretizations of QFTs that LGTs are, and these more recent QCA discretizations.

Let us comment on what happens to the maximum speed in LGTs.
In continuous-time, i.e.\ Hamiltonian LGTs, the evolution operator is not ultralocal.
%
%More generally, CTQWs are not ultralocal evolutions.
%
Indeed, for the dynamics described by a lattice Hamiltonian to be non-trivial, this Hamiltonian must be non- block-diagonal in position space, otherwise there is no interblock dynamics;
but the evolution operator corresponding to a non- block-diagonal Hamiltonian is generically not ultralocal, i.e., from one instant to another, there is always a non-zero probability to have moved arbitrarily far from the starting position\footnote{Notice that, out of a certain light cone, i.e., for big enough distances $x$ and small enough times $t$ such that $x > V_{\text{LB}} t$, where $V_{\text{LB}}$ is the Lieb-Robinson (LB) upper bound velocity on the group velocity \cite{Lieb1972}, this probability is, although generically non-vanishing, exponentially bounded with distance, independently of the state. This result is generic to near-neighbors Hamiltonian, i.e., Hamiltonians with finite-support interactions and defined on lattices, and it applies, more generally, to correlations between any two observables separated by some distance. The LB bound acts in practice, i.e., up to negligible errors, as a strict upper bound, i.e.\ causally decorrelates distant parts of the system, since an exponential suppression is a very strong one. However, the LB bound is, on the contrary to the speed of light, not universal: it depends on the considered Hamiltonian, that is, both on the structure of the lattice and on the form of the near-neighbors interaction.}.
Let us now speak about the discrete-time versions of LGTs. 
These being formulated, for technical reasons, in Euclidean spacetime, their unitarity is not manifest.
Unitarity can be proven in certain cases, e.g., for the (1+1)D Wilson \cite{Creutz1977} and staggered \cite{Sharatchandra1981} LGT models in Euclidean discrete-time, the continuous-time version of which we start from, further down.
However, generalizing these models to higher dimensions and number of flavors can lead to difficulties in ensuring unitarity \cite{Hernandez2011,Smit1991}, while QCA discretizations are by construction manifestly unitary.
Let us now come back to the question of the ultralocality of the evolution.
The transfer matrices, that is, the Euclidean versions of the evolution operators,
%, exhibited in the two previsoulsy-cited works  \cite{Creutz1977,Sharatchandra1981},
are built by exponentiating the terms of the spacetime-discretized action in a way which ensures the positive definiteness of the transfer matrix, i.e., the unitarity of the evolution.
This criterion does not forbid the exponentiation of non- block-diagonal matrices, and it is no surprise that the obtained transfer matrices, and so the associated evolution operators, are \emph{not} ultralocal.
The digitization performed in Martinez et al.'s experiment does not yield an ultralocal evolution either.
In contrast, QCA are manifestly ultralocal by construction.

The present work is aimed at shedding light on conceptual and technical relationships which exist between LGTs and QCA discretizations of QFTs.
%
%More precisely, it is shown, in (1+1)D spacetime and with a single-particle Hamiltonian-LGT model (as opposed to those describe by actions), that the ultralocal discrete-time evolution resulting from the above-mentioned even-odd exponentiation, is nothing but a DTQW.
%
The main result is that DTQWs can be used to digitize fermionic models of continuous-time LGTs, in such a way that the evolution operator is not only manifestly unitary, but also ultralocal.
Our discrete-time ultralocal scheme is not chiral symmetric, neither (i) in the standard multicomponent-wavefunction picture, nor (ii) in the staggered-fermions picture, where the wavefunction is scalar and the chiral symmetry corresponds to a translational invariance of the model. While (i) could be satisfied at the price of introducing fermion doubling, which our model avoids, (ii) is unavoidable for an ultralocal unitary evolution, in virtue of Meyer's 1996 no-go lemma stating that no non-trivial one-dimensional scalar quantum cellular automaton can be translationally invariant \cite{Meyer96a}.
%
%More precisely, the example we describe suggests that it may be difficult to retain a lattice chiral symmetry of the type that staggered fermions have, if one requires the ultralocality of the evolution operator, at least if one sticks to time-discretization method followed in the present work.
%

\paragraph*{{Left-right spatial discretization of the (1+1)D Dirac equation: motivation, definition and properties.}} \label{sec:1}

The (1+1)D Dirac equation reads
\(
i \partial_t \Psi= \mathcal{H}_0(-i \partial_x) \Psi,
\)
having introduced (i) a two-component Dirac wavefunction,
\( \Psi_{(t,x)} = \langle x | \Psi_{(t)} \rangle = (
\psi^L_{(t,x)},
\psi^R_{(t,x)} 
)^{\top},
\)
\(\top\) denoting the transposition, with superscripts $L$ and $R$ for `left' and `right', explained after Eqs.\ (\ref{eq:CT_lim}), and (ii) the free Dirac Hamiltonian,
\(
\mathcal{H}_0(-i \partial_x) = \alpha^1 (-i \partial_x) +  m \alpha^0,
\)
with mass $m$ and alpha matrices
\(
\alpha^0 = \sigma^3
\)
and
\(
\alpha^1 = \sigma^1,
\)
where
\(\sigma^n\) is the \(n\)th Pauli matrix. 
We introduce a 1D spatial lattice \((x_p=pa)_{p \in \mathbb{Z}}\) with lattice spacing \(a\).
The so-called \emph{naive} spatial discretization of the above Dirac equation is obtained by replacing \(\partial_x\)  by a symmetric finite difference:
\(
i \dot{\Psi}_{(t,x)}= \alpha^1 [ -i \, (\Psi_{(t, x+a)} - \Psi_{(t,x-a)})/(2a) ] + m \alpha^0 \Psi_{(t,x)}.
\)
This way of discretizing is known to suffer from the fermion-doubling problem \cite{Susskind77a, book_Rothe},
which comes from the use of finite differences defined over two lattice spacings rather than a single one \cite{book_Rothe}. 
The mere replacement of the symmetric finite difference by an asymmetric one breaks the Hermiticity of the Hamiltonian, and leads to renormalization issues \cite{book_Rothe}. 
That being said, it is possible to preserve Hermiticity while sticking to asymmetric finite differences, by the use of, say, a left (resp.\ right) finite difference for the upper (resp.\ lower) component of the Dirac wavefunction.
Such a discretization can be written as
\(
i \ket{\dot{\Psi}_{(t)}} = \hat{H} \ket{\Psi_{(t)}},
\)
where \(\hat{H}\) will be called the \emph{left-right Hamiltonian}, and reads
\(
\hat{H} = \sum_p \hat{H}_p,
\)
each single-site being the sum of two terms,
\(
\hat{H}_p =   \hat{H}_p^{\text{m}}  + \hat{H}_p^{\text{t}},
\)
a mass term,
\(
\hat{H}_p^{\text{m}} = m \alpha^0 \ket p \! \! \bra{p},
\)
and the announced left-right transport term,
{\normalsize \(
\hat{H}_p^{\text{t}} = (-i /a)\  \text{antidiag}(
\ket{p} \! \! \bra{p} -  \ket{p+1} \! \! \bra{p}, \\
\ket p \! \! \bra{p+1} -  \ket{p} \! \! \bra{p})
\)},
where the first `coefficient' is the upper-right one, and \(\ket{p} = \ket{x_p}\).
Notice that \(\hat{H}\) is (i) translationally invariant and (ii) of near-neighbors type.
The equations of motion induced by the left-right Hamiltonian read
\begin{subequations} \label{eq:CT_lim}
\begin{align}
i \dot{\psi}^{L}_{p} &= \frac{-i}{a} \left( \psi^{R}_{p} - \psi^{R}_{p-1} \right) + m \psi^{L}_{p} \\
i \dot{\psi}^{R}_{p} &= \frac{-i}{a} \left( \psi^{L}_{p+1} -  {\psi}^{L}_{p} \right) - m {\psi}^{R}_{p} \, ,
\end{align}
\end{subequations}
with \(\Psi_p = \Psi_{(x_p)}\), and where we have omitted the time variable to lighten notations.
Now, look at the first equation: \(\psi^L_p\) is `fed', i.e.\ \(\dot{\psi}^L_p\) is determined, by the knowledge of \(\psi^R_p\), which is at the same location, and of \(\psi^R_{p-1}\), which is on the left of \(\psi^L_p\), so that we may say that \(\psi^R_{p}\) feeds \emph{to the right},  hence the superscript \(R\), and similarly for \(L\) (second equation). 

As mentioned before, the left-right discretization avoids fermion doubling. 
Now, there exists a well-known solution to remove fermion doubling in naive discretizations, namely, that of Wilson \cite{WK1974}, known as Wilson fermions, which consists of adding, to the standard naive discretization, a mass term of the Schrödinger type (i.e., a lattice Laplacian), called Wilson term.
The effect of the Wilson term on the dispersion relation is to raise the energy on the edges of the Brillouin zone in order to remove the unwanted extra poles in the propagator, and this can be done by a tunable amount \cite{Susskind77a}.
The Hamiltonian corresponding to this model, that we may call Wilson's Hamiltonian, is \cite{Susskind77a}
\( 
\hat{H}_{\text{W}}^{(r)} = \sum_p ( \hat{H}_{\text{W}}^{(r)} )_p,
\)
with 
\(
( \hat{H}_{\text{W}}^{(r)} )_p = ( \hat{H}_{\text{n}})_p + (\hat{H}_{\text{S}}^{(r)})_p,
\)
where \(\text{W}\) is for `Wilson', \(\text{n}\) for `naive', \(\text{S}\) for `Schrödinger', and \(r \in \mathbb{R}\) is Wilson's parameter. The naive Hamiltonian is given by
\(
(\hat{H}_{\text{n}})_p = (\hat{H}^{\text{m}}_{\text{n}})_p  + (\hat{H}_{\text{n}}^{\text{t}})_p,
\)
where  we write the mass term in a different representation from that above,
\(
(\hat{H}_{\text{n}}^{\text{m}})_p = m (-\sigma^2) \ket p \! \! \bra{p},
\)
and where the naive transport term is given by
\(
(\hat{H}_{\text{n}}^{\text{t}})_p = \frac{-i}{2a} \alpha^1 \big(  \ket{p} \! \! \bra{p+1} -  \ket{p+1} \! \! \bra{p}  \big).
\)
Finally, Wilson's term is given by
\(
(\hat{H}_{\text{S}}^{(r)})_p = \alpha^0 \frac{r}{2a}  \big(   2 \ket{p} \! \! \bra{p} - \ket{p} \! \! \bra{p+1} -  \ket{p+1} \! \! \bra{p}  \big).
\)
The choice \(r=1\) is the most popular one, and we will stick to it unless otherwise mentioned.
Now, it turns out that the left-right Hamiltonian is unitarily equivalent to Wilson's Hamiltonian with \(r=1\), via a certain rotation in the internal Hilbert space, that is,
\(
B \hat{H}_p B^{\dag} = (\hat{H}_{\text{W}}^{(r=1)})_p,
\)
where the unitary matrix is
\(
B = \exp(-i\sigma^1\frac{\pi}{4}).
\)
Note that, in order to lighten the writing, we have omitted the identity tensor factor of the position Hilbert space, and will do so from now on in similar cases.

It is well known that adding a Wilson term such as that above breaks the \emph{axial-U(1)}, also called \emph{chiral} symmetry, of the naive lattice massless Dirac Hamiltonian, i.e.\ the massless $\hat{H}_{\text{W}}^{(r\neq 0)}$ does not commute with \(\gamma^5\), which equals \(\pm i^{d/2+1}\gamma^0...\gamma^{d-1}\) in even spacetime dimensions \(d\), that is, in the present case, say \(\gamma^5=+\sigma^1\), which amounts to the exchange of the upper and lower components of the Dirac wavefunction.
The left-right Hamiltonian also breaks chiral symmetry\footnote{It is actually a general result, known as the Nielsen-Ninomiya no-go theorem, that no Hamiltonian that respects the usual Hermiticity, locality (i.e.\ near-neighbors structure) and translation-invariance conditions, such as \(\hat{H}\) or \(\hat{H}_{\text{W}}^{(r)}\), can avoid fermion doubling if it does not break chiral symmetry.}.
A well-known solution to restablish, up to a modification of the Hilbert space, chiral symmetry (without, of course, reintroducing the doubling problem), is to work with the so-called \emph{staggered} formulation of LGT \cite{KogSuss75a, book_Rothe}.
The idea is to distribute the internal components of the Dirac wavefunction over different lattice sites.
%
%Now, the left-right Hamiltonian that we have introduced above is a non-staggered Hamiltonian, which is particularly simply related to the staggered (1+1)D Hamiltonian, defined as follows.
%
Consider Eqs.\ (\ref{eq:CT_lim}).
Take every lower component \(\psi^R_p\), and shift it spatially by \(a/2\), i.e.\ position it at \(x=pa + a/2\) (this is the only operation to be performed).
This induces a new lattice \((x_n=n a/2)_{n \in \mathbb{Z}}\) of spacing \(a'=a/2\) which is filled, at even sites \(n=2p\), with \(\psi^{\text{e}}_{2p} \equiv \psi^L_{p}\), and, at odd sites \(n=2p+1\), with \(\psi^{\text{o}}_{2p+1} \equiv \psi^R_{p}\).
We can thus define a single-component wavefunction \(\phi\) such that
\(
\phi_n = \psi^{\text{e}}_{n} \ \text{for $n$ even}, \ \text{and} \
\psi^{\text{o}}_{n} \ \text{for $n$ odd}.
\)
In terms of \(\phi\), Eqs.\ (\ref{eq:CT_lim}) can be recast as a single equation,
\(
i \dot{\phi}_{n} = \frac{-i}{2a'} \left( \phi_{n+1} - \phi_{n-1} \right) + m (-1)^n \phi_{n},
\)
that is,
\(
i \ket{\dot{\phi}} = \hat{H}_{\text{stag.}} \ket{\phi},
\)
where 
\(
\hat{H}_{\text{stag.}} = \sum_n (\hat{H}_{\text{stag.}})_n,
\)
with 
\(
(\hat{H}_{\text{stag.}})_n =  (\hat{H}_{\text{stag.}}^{\text{m}})_n + (\hat{H}_{\text{stag.}}^{\text{t}})_n, \ (\hat{H}_{\text{stag.}}^{\text{m}})_n = m (-1)^n \ket{n} \! \! \bra{n}, \
(\hat{H}_{\text{stag.}}^{\text{t}})_n = \frac{-i}{2a'}  \big(  \ket{n} \! \! \bra{n+1} -  \ket{n+1} \! \! \bra{n}  \big).
\)
%
%Before discussing the recovery of a remnant of chiral symmetry in the staggered discretization, notice the following.
%
%First, a Hamiltonian has chiral symmetry, i.e.\ commutes with $\exp(i\beta\gamma^5)$, if and only if it commutes with $\gamma^5$.
%
%Moreover, in 1+1 dimensions and with our Clifford-algebra representation choice, one can choose $\gamma^5 = - ( i^{(2/2)+1} \gamma^0 \gamma^1 ) = \alpha^1 = \sigma^1$, so that commuting with $\gamma^5$ corresponds to an invariance under the exchange of the upper and lower components of the Dirac wavefunction.
%
Now, the massless staggered Hamiltonian is invariant, not only by two-site translations on the lattice, which simply corresponds to the translational invariance of the original non-staggered Hamiltonian, but also by single-site translations on that lattice, which corresponds to the exchange of the original upper and lower components, that is, to the \(\gamma^5\) invariance of the massless continuum Dirac Hamiltonian. 
This is how the staggering restablishes a remnant\footnote{`Remnant' in the sense that the symmetry is structurally different: it is not a symmetry in an internal Hilbert space (there is no such space anymore), but in the external one.} of chiral symmetry.

To sum up, although both the left-right discretization and the staggered one avoid fermion doubling and might seem equivalent at first sight, in the sense that one can, picturingly speaking, go from the left-right to the staggered one by a graphical rotation of \(\pi/2\) of all column vectors \((\psi^L_p,\psi^R_p)^{\top}\), \(p \in \mathbb Z\), in order to position the lower components on different lattice sites, this graphical rotation, i.e., the transformation of internal degrees of freedom into external ones, is actually crucial to recover a remnant of chiral symmetry.

\begin{figure}
\includegraphics[width=8.7cm]{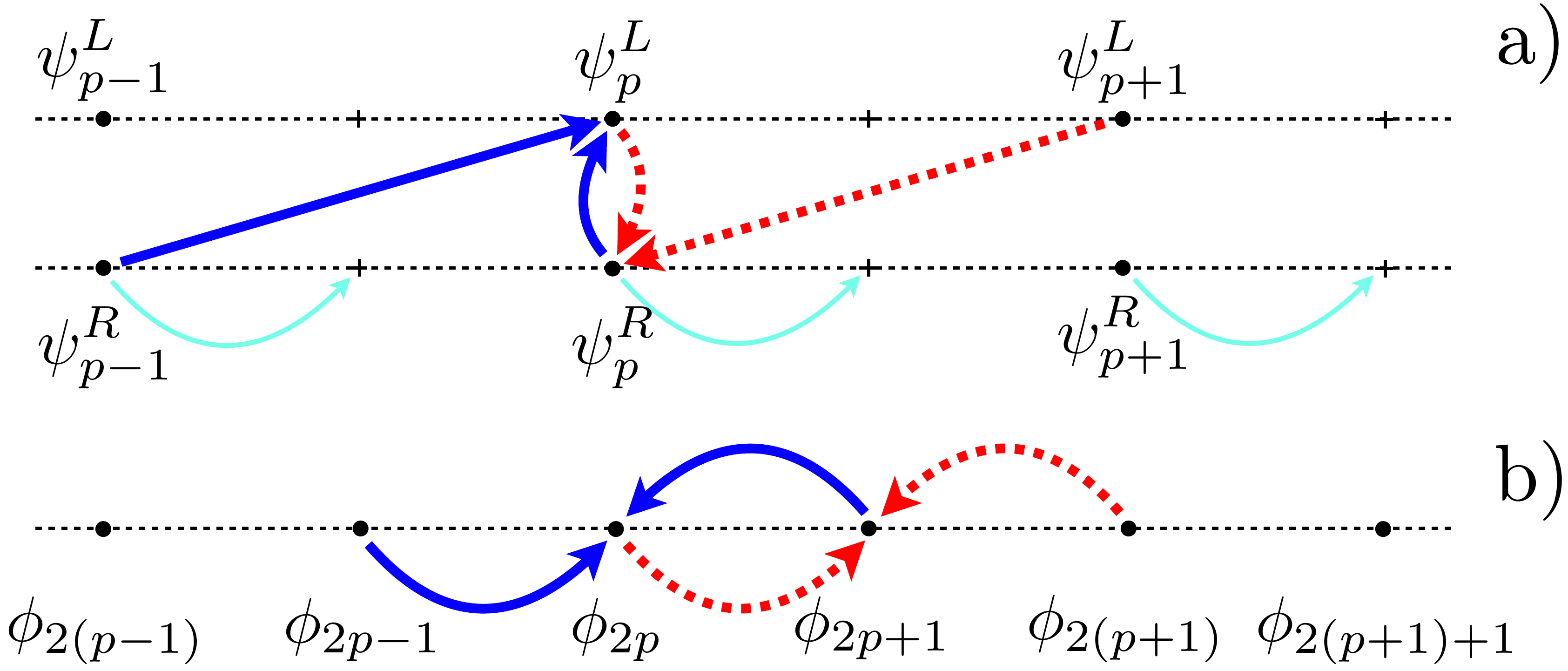}
\caption{Comparison between the continuous-time a) left-right and b) staggered dynamics. The thick, solid blue (resp.\ dashed red) arrows indicate how the \(\psi^R_p\)'s or \(\phi_{2p+1}\)'s (resp.\  \(\psi^L\)'s or \(\phi_{2p}\)'s feed the \(\psi^L_p\)'s or \(\phi_{2p}\)'s (resp.\  \(\psi^R\)'s or \(\phi_{2p+1}\)'s). The thin, light-blue arrows indicate a possible way of performing the staggering starting from the left-right discretization. One can see graphically that the staggered scheme restablishes chiral symmetry.}
\end{figure}

\paragraph*{{Ultralocal unitary digitization of the spatially-discretized (1+1)D Dirac equation via DTQW.}}

We are interested in the time discretization of the continuous-time unitary dynamics generated by the left-right Hamiltonian, which is non- block-diagonal in position space.
Now, the naive time discretization of non- block-diagonal lattice Hamiltonians, that is, considering the continuous-time evolution but through a stroboscope of period \(\Delta t\), leads to an evolution between two discrete-time instants which is, generically, \emph{not} ultralocal, because the exponential of a non- block-diagonal Hamiltonian is not ultralocal.
One may say, in the present case of a near-neighbors Hamiltonian, that the naive time discretization leads to a `loss of ultralocality', in the sense that the near-neighbors structure of the Hamiltonian is the most local continuous-time dynamics that one can conceive on a lattice, which may also be seen, in terms of the evolution operator, as an ultralocality, but `at constant time'.
The ultralocality of the evolution cannot be naively `restored', after time discretizing, by truncation of the exponential series, because this breaks unitarity.
A well-known trick to `restore' ultralocality in the time discretization of a near-neighbors Hamiltonian, is to split the Hamiltonian into block-diagonal parts, and then use the Trotter-Suzuki approximation to build an appropriate ultralocal one-time-step evolution operator \cite{Patel2005}. 
Let us do so for \(\hat{H}\).

The mass term is irrelevant in this discussion, since it can be  time-discretized naively within the Trotter-Suzuki scheme because it is diagonal in position space. 
Let us focus on the transport Hamiltonian \(\hat{H}^{\text{t}}\). 
One can write
\(
\hat{H}^{\text{t}} = \hat{H}^{\text{on}} +  \hat{H}^{\text{int.}},
\)
where, for $\text{i} = \text{on}, \text{int.}$,
\(
\hat{H}^{\text{i}} = \sum_p \hat{H}^{\text{i}}_p,
\)
with the on-site and the intersite single-site terms, respectively given by
\(
\hat{H}^{\text{on}}_p = \frac{1}{a} \sigma^2  \ket{p} \! \! \bra{p},
\)
and
\(
\hat{H}^{\text{int.}}_p = \frac{1}{a} \big[ -i \sigma^{+} \ket{p} \! \! \bra{p+1} + \text{H.c.} \big],
\)
where
\(
\sigma^+ = (\sigma^1 + i \sigma^2)/2.
\)
\(\hat{H}^{\text{on}}\) is manifestly block diagonal in position space. 
What about \(\hat{H}^{\text{int.}}\)? 
To visualize the situation, let us explicitly write the matrix representation \(\mathbf{H}^{\text{t}}\) of this transport Hamiltonian, \(\hat{H}^{\text{t}}\), in the \(LR\)-position basis, \(\big( \ket p \otimes \ket L,  \ket p \otimes \ket R  \big)_{p \in \mathbb{Z}}\).
\iffalse
, whose column-vector representation is
\(
\big( | p) \otimes |L),  |p) \otimes |R)  \big)_{p \in \mathbb{Z}},
\)
where \(|v)\) is the canonical column-vector representation of the ket \(\ket v\), i.e.\
\(
|L)  = (
1, 0)^{\top}, 
|R)  =
(0, 1)^{\top},
|p)  = (
\hdots, 0, 1, 0, \hdots)^{\top},
\)
the \(1\) being located on the \(p\)th row.
%
Let us write \(\mathbf{H}^{\text{t}}\) in the Kronecker-product basis defined above. 
%
We order the columns (resp.\ rows) with growing \(p\)'s from left to right (resp.\ top to bottom), and each \(p\) is associated to two consecutive columns (resp.\ rows), labelled \(pL\) and \(pR\) from left to right (resp.\ top to bottom). 
%
\fi
This yields, from, say, \(p-1\) to \(p+1\),
{\footnotesize
\begin{align}\label{eq:Ht}
\mathbf{H}^{\text{t}} &=  \frac{-i}{a} \left[
\begin{array}{*{9}{@{}C{1cm}@{}}}
 \cdot & \fbox{$1$} & \cdot & \cdot &\cdot& \cdot \\
\fbox{$-1$} & \cdot & 1 & \cdot &\cdot& \cdot  \\
\cdot & -1 & \cdot & \fbox{$1$} &\cdot & \cdot  \\
\cdot & \cdot &\fbox{$-1$} & \cdot & 1 & \cdot  \\
\cdot &  \cdot &\cdot &-1&  \cdot & \fbox{$1$} \\
\cdot & \cdot & \cdot & \cdot &\fbox{$-1$}& \cdot
\end{array} 
\right] \, .
\end{align}
}
\noindent
For a good visualization, we have only written a \(6 \times 6\) matrix, but the equality stands for the infinite-dimensional matrix. 
The dots stand for zeros. 
The boxed \(1\)'s and \((-1)\)'s correspond to \(\hat{H}^{\text{on}}\), and the others to \(\hat{H}^{\text{int.}}\). 
We notice the above-mentioned manifest block-diagonality of \(\hat{H}^{\text{on}}\) in position space. 
Now, what this matrix representation `reveals', is that \(\hat{H}^{\text{int.}}\) is also block-diagonal in the full Hilbert space, obviously not in the position basis \((\ket{p})_{p\in\mathbb{Z}}\), that we shall call non-staggered position basis, but in the complete, i.e.\ \(LR\)-position basis, or, via the correspondence \(Lp \rightarrow 2p\) and \(Rp \rightarrow 2p+1\), in the staggered position basis, \((\ket{n})_{n\in\mathbb{Z}}\). Indeed, this correspondence yields, not only the following identification,
\(
\mathbf{H}^{\text{t}} = \mathbf{H}^{\text{t}}_{\text{stag.}},
\)
where \(\mathbf{H}^{\text{t}}_{\text{stag.}}\) is the matrix representation of the massless staggered Hamiltonian, but also that the on-inter splitting of \(\hat{H}\) is nothing but a standard even-odd splitting of \(\hat{H}_{\text{stag.}}\), similar to that performed in Ref.\ \cite{Patel2005}, that is,
\(
\mathbf{H}^{\text{on}} = \mathbf{H}^{\text{e}}, 
\)
and
\(
\mathbf{H}^{\text{int.}} = \mathbf{H}^{\text{o}},
\)
where
\(
\hat{H}_{\text{stag.}} =  \hat{H}^{\text{e}} +  \hat{H}^{\text{o}},
\)
with the even and odd parts respectively given by
\(
\hat{H}^{\text{e}}  = \sum_p (\hat{H}^{\text{t}}_{\text{stag.}})_{2p} 
\) 
and
\(
\hat{H}^{\text{o}}  = \sum_p (\hat{H}^{\text{t}}_{\text{stag.}})_{2p+1}.
\)

Let us introduce a discrete time coordinate \(j\in\mathbb{N}\), such that \(\Psi_{j,p} = \Psi_{(t_j,x_p)}\), where \(t_j=j\Delta t\), and \(\Delta t\geq 0\) is the time step.
Now that we have split \(\mathbf{H}^{\text{t}} = \mathbf{H}^{\text{t}}_{\text{stag.}}\) into two block-diagonal parts, we can build the desired ultralocal time-discretized evolution from time \(j\) to time \(j+1\), that we write
\(
\ket{\Psi_{j+1}} = \mathbf{U} \ket{\Psi_{j}} + O(\Delta t) 
\)
-- where, to lighten the writing, we have simply used the notation \(\ket{\Psi_{j}}\) for the associated matrix representation --, by defining
\(
\mathbf{U} \equiv \mathbf{U}^{\text{m}}  \mathbf{U}^{\text{t}},
\)
with
\(
\mathbf{U}^{\text{t}} \equiv \mathbf{U}^{\text{on}} \mathbf{U}^{\text{int.}} \, ,
\)
where, for 
\(\text{i} = \text{m}, \text{on}, \text{int.}\),
\(
\mathbf{U}^{\text{i}} \equiv \exp(-i \Delta t \, \mathbf{H}^{\text{i}})  \, .
\)
Note that we have interpreted \(\mathbf{U}\) as an evolution operator in the left-right picture, but we can also interpret it in the staggered picture.
Now, we have seen above that both the mass and the on-site Hamiltonians are block-diagonal in the non-staggered position basis, so that one can perform a block exponentiation.
Since the involved blocks are Pauli matrices, which square to the identity, their exponentiation is straightforward (use, e.g., the power-series representation of the exponential). 
Moreover, we have also seen that the intersite Hamiltonian is actually the same as the on-site one but shifted by one lattice site in the staggered position basis. 
In the end, we thus obtain  \(\mathbf{U}^{\text{m}} = \text{diag} (\mu,\mu^{\ast},\mu,\mu^{\ast},\mu,\mu^{\ast})\), where \(\mu = \exp(-i \Delta t \, m)\), the \(\ast\) denotes complex conjugation, and
{\small
\begin{align}
\mathbf{U}^{\text{on}} \! = \! \! \left[
\begin{array}{*{9}{@{}C{0.4cm}@{}}}
c & -s & \cdot &\cdot& \cdot & \cdot \\
s & c & \cdot & \cdot & \cdot & \cdot \\
\cdot & \cdot & c & -s & \cdot & \cdot  \\
\cdot & \cdot & s & c & \cdot & \cdot \\
\cdot &\cdot& \cdot & \cdot & c & -s \\
\cdot &\cdot& \cdot & \cdot &  s & c 
\end{array}
\right] \! , \
\mathbf{U}^{\text{int.}} \!=  \! \! \left[ 
\begin{array}{*{9}{@{}C{0.4cm}@{}}}
c & \cdot & \cdot & \cdot &\cdot& \cdot \\
\cdot & c & -s & \cdot &\cdot& \cdot  \\
\cdot & s& c & \cdot &\cdot & \cdot  \\
\cdot & \cdot &\cdot &c & -s & \cdot  \\
\cdot &  \cdot &\cdot &s& c & \cdot  \\
\cdot & \cdot & \cdot & \cdot &\cdot & c
\end{array}
\right] \!, \! \!  \label{eq:Uextt}
\end{align}
}
where
\(
c = \cos \delta
\) 
and
\(
s = \sin \delta,
\)
with
\(
\delta = \Delta t/a.
\)
A straightforward computation delivers the product
{\small
\begin{align} \label{eq:Utbis}
\mathbf{U}^{\text{t}} =  \left[ 
\begin{array}{*{9}{@{}C{1cm}@{}}}
c ^2 & - sc & s^2 & \cdot &\cdot& \cdot \\
 sc & c^2 & -sc & \cdot &\cdot& \cdot  \\
\cdot & sc& c^2 &  - sc & s^2 &  \cdot \\
\cdot &  s^2 & sc& c^2 &  - sc & \cdot    \\
\cdot &\cdot& \cdot &  sc & c ^2 & -sc \\
 \cdot &\cdot& \cdot  & s^2 & sc & c^2
\end{array}
\right] \, ,
\end{align}}
which is translationally invariant in the non-staggered position basis. 
The discrete-time evolution through \(\mathbf{U}\) reads
\(\Psi_{j+1,p} = \langle p| \mathbf{U} |\Psi_j \rangle,
\)
that is, explicitly,
{\small
\begin{align} \label{eq:W}
\psi_{j+1,p}^{L} &= e^{-i\Delta t \, m} \left(  sc \, \psi_{p-1}^{R} + c^2 \psi^{L}_{p} - sc \, \psi_{p}^{R} + s^2 \psi_{p+1}^{L} \right)  \nonumber \\
\psi_{j+1,p}^{R} &=  e^{+i\Delta t \, m} \left( s^2 \psi_{p-1}^{R} + sc \, \psi_{p}^{L} + c^2 \psi^{R}_{p} - sc \, \psi_{p+1}^{L} \right)  \, .
\end{align}}
Recall that, in the limit $\Delta t \rightarrow 0$, these equations coincide with the dynamics of continuous-time LGT fermions, which in turn coincides, in the limit of a lattice spacing $a \rightarrow 0$, with standard Dirac dynamics.

The ultralocal transport evolution operator, \(\mathbf{U}^{\text{t}}\), that we have built thanks to the even-odd splitting of the staggered Hamiltonian \(\mathbf{H}^{\text{stag.}}\), has `lost' the single-site translation invariance of \(\mathbf{H}^{\text{stag.}}\).
This is the price to pay for this even-odd digitization, which renders the discrete-time scheme ultralocal: indeed, Meyer's 1996 no-go lemma states that no non-trivial one-dimensional scalar ultralocal unitary evolution, that is, quantum cellular automaton, can be translationally invariant \cite{Meyer96a}. 
Hence, \emph{there is no more remnant of chiral symmetry in} \(\mathbf{U}^{\text{t}}\). 
This makes the left-right picture more relevant than the staggered one in a discrete-time framework, since the translational invariance is only realized in the former.
Let us now state one of the main points of the present work.
It turns out that the left-right- picture
%\footnote{What follows in the main text can also be written in the staggered picture, by replacing \(L\) (resp.\ \(R\)) by `even' (resp.\ `odd') and by using the the staggered-picture lattice, but this demands to be able to realize, on this lattice, (two-site) translations of even-site (resp.\ odd-site) components without translating the odd-site (resp. even-site) ones. In other words, if such translations can be realized, no single-site translations are needed to evolve the walker on the staggered-picture lattice, and the suggested procedure is conceptually equivalent to the left-right picture since it naturally introduces an even-odd internal Hilbert space. Such an even-odd picture with even-odd internal degree of freedom should thus simply be seen as a possible instance of the left-right picture, with possible experimental interest\label{footnote:EO}})
\(\mathbf{U}^{\text{t}}\) in Eq.\ (\ref{eq:Utbis}) can be written as a DTQW of the type introduced by Strauch to establish a connection between DTQWs and continuous-time quantum walks \cite{Strauch06b}, namely,
\begin{align} \label{eq:compact}
\mathbf{U}^{\text{t}}= C(-\theta) S^R_{\mathbf{k}} C(\theta)  S^L_{\mathbf{k}} \, ,
\end{align}
where we have  introduced (i) a coin operation
\(
C(\theta) = \exp(-i\sigma^2 \theta/2),
\)
where 
\(
\theta = \pi - 2 \delta,
\)
and (ii) left and right internal-state- dependent shifts,
\(
S^L_{\mathbf{k}} = \text{diag}(
e^{i \mathbf{k}},1)\)
and
\(
S^R_{\mathbf{k}} = \text{diag}(
1, e^{-i \mathbf{k}}),
\)
where \(\mathbf{k}\) is the quasimomentum operator associated to the matrix representation of the position basis \((\ket{p})_{p\in\mathbb{Z}}\). 
Note that \(\mathbf{U}^{\text{t}}\) is invariant under the exchange of \(S^R_{\mathbf{k}}\) and \(S^L_{\mathbf{k}}\), i.e.\ \(\mathbf{U}^{\text{t}}= C(-\theta) S^L_{\mathbf{k}} C(\theta)  S^R_{\mathbf{k}}\), which simply indicates that at each time step, one can first shift either the left movers or the right movers. 
%
%Eventually, note the following. One can easily check that the left-right Hamiltonian, $\hat{H}=\hat{H}^{\text{m}} + \hat{H}^{\text{on}} + \hat{H}^{\text{int.}}$, has CPT symmetry, but neither P, nor CP. All three $\hat{H}^{\text{m}}$, $\hat{H}^{\text{on}}$, and $\hat{H}^{\text{int.}}$, also have CPT symmetry, so that so do their respective evolution operators, as well as the product of these, so that $\mathbf{U}$ has CPT symmetry.
%Let us mention that those DTQWs (such as $\mathbf{U}^{\text{t}}$) which cannot be built solely out of internal-state- dependent shifts which do not leave some internal components of the wavefunction at the same site, are sometimes called split-step DTQWs.

In the Supplemental Material, we extend our time-discretization method to Wilson's Hamiltonian in the original internal-space representation, which allows for any choice of Wilson's parameter, \(r\). 
We also U(1)-gauge our DTQWs; the lattice gauge transformations involve, notably, the \emph{standard} finite differences used in LGT, instead of the more complicated ones used in Ref.\ \cite{AD16}.
We suggest a gauge-invariant quantity on the spacetime lattice and a classical on-shell dynamics (Maxwell's equations) for it, more appropriate than that of Ref.\ \cite{AD16}.

\iffalse
In the Supplemental Material, we show (A)  that the naive spatial discretization of the (1+1)D Dirac equation and (B) the standard additional Wilson term can also be digitized ultralocally (and unitarily) with DTQWs, via the same method or a very similar one, respectively. 
%
We also (C) give the precise connection between the DTQWs introduced in the present work and Strauch's original one.
%
We (D) perform the standard even-odd, non- DTQW-based digitization of the (1+1)D Dirac equation and show that it is unitarily mapped to the DTQW-based one, but that this mapping is \emph{not} of near-neighbors type. 
%
This implies, in particular, that it is, in the number of spatial-lattice sites, quadratically costly to go from the non- DTQW-based digitization to the DTQW-based one or vice versa -- while this would be linearly costly if the mapping was of near-neighbors type. 
%
Also (E), we show how to include a U(1) gauge-field coupling in our scheme, Eq.\ (\ref{eq:compact}), which gives back the known models of Hamiltonian LGT and continuum field theory when taking, respectively, the continuous-time limit and an additional continuous-space limit.
%
This U(1)-gauged scheme is gauge invariant on the spacetime lattice, and the gauge transformations involve, notably, the \emph{standard} finite differences used in LGT, instead of the more complicated ones used in Ref.\ \cite{AD16}.
%
We suggest a gauge-invariant quantity on the spacetime lattice and a classical on-shell dynamics (Maxwell's equations) for it, more appropriate than that of Ref.\ \cite{AD16}. 
\fi

\paragraph*{{Conclusion and perspectives.}}

Enforcing the ultralocality of the evolution operator leads to a loss of the staggered-model chiral symmetry, and this is unavoidable in virtue of Meyer's 1996 no-go lemma stating that no non-trivial one-dimensional scalar quantum cellular automaton can be translationally invariant.
Sharatchandra et al.'s staggered discrete-time scheme is (unitary and) chiral in the staggered sense, i.e. translationally invariant, but not ultralocal \cite{Sharatchandra1981}, while our discrete-time scheme is (unitary and) ultralocal, but not translationally invariant.
%
%By extrapolating naively from the present work, it seems, generally, difficult to impose both the requirement of lattice chiral symmetry and that of ultralocality of the evolution operator.%
Can one find a DTQW i.e.\, an ultralocal discrete-time evolution, in 1+$d$ dimensions with $d \geq 3 $ (in $d=2$ there is no notion of chirality), (i) having staggered fermions as a continuous-time limit, and (ii) which preserves, in discrete time, the lattice chiral symmetry of staggered fermions?

Our discrete-time scheme is ultralocal, so that it shares with continuum QFTs the property that the particle's speed is upper bounded.
However, we can only recover the appropriate continuum-limit equations by first performing a continuous-time limit, which is a non-relativistic limit in which the speed of light goes to infinity.
In other words, setting $ \Delta x / \Delta t = c$, a constant, makes it impossible to derive either a continuous-time limit or a continuous-spacetime one.
This forbids the straightforward identification of the maximum speed of the discrete-time model as a discrete-time counterpart of the speed of light of the Dirac equation in the continuum, a question we hope to solve in future work.
Recall, indeed, that this difficulty contrasts with the standard DTQW-discretized Dirac equation, in which the discrete-spacetime counterpart of the continuum speed of light is precisely simply the ratio $\Delta x / \Delta t$, which can be set constant without forbidding the continuous-spacetime limit to be performed.

\paragraph*{{Aknowledgments.}}
P.\ Arnault thanks Michael Creutz, Erez Zohar, Mari Carmen\ Ba\~nuls, Pilar Hern{\'a}ndez, and Andrea Alberti, for their insights and suggestions.
%
%Part of this work has been done at LERMA, UMR 8112, UPMC and Observatoire de Paris, 4 place Jussieu, 75005 Paris, France. 
%
This work has been supported by the Spanish Ministerio de Educaci{\'o}n e Innovaci{\'o}n, MINECO-FEDER project FPA2017-84543-P, SEV-2014-0398 and Generalitat Valenciana grant GVPROMETEOII2014-087.

%\bibliography{bibli.bib}


\newpage

\appendix

\setcounter{footnote}{0}

{\small
\begin{widetext}
\begin{center}
 {\large {\bfseries Supplemental Material } \\
about \\
{\bfseries  Discrete-time quantum walks as fermions of lattice gauge theory}}
\end{center}

%\iffalse
%\setsinglecolumn
%\begin{center}
{This Supplemental Material contains, most notably, the following appendices. In Apps.\ \ref{app:naive_case} and \ref{app:Wilson_term} we show, respectively, that (i) the naive spatial discretization of the (1+1)D Dirac equation and (ii) the standard additional Wilson term can also be digitized unitarily and ultralocally with DTQWs, via (i) the same method or (ii) a very similar one, respectively. 
In App.\ \ref{app:Strauch}, we give the precise link between the DTQWs introduced in the present work and Strauch's original one.
In App.\ \ref{app:non-chiral}, we perform the standard even-odd, non- DTQW-based digitization of the (1+1)D Dirac equation, and show that it is unitarily mapped to the DTQW-based one, but that this mapping is \emph{not} of near-neighbors type, i.e.\ it has matrix elements arbitrarily far from the diagonal.
This implies, in particular, that it is quadratically costly, in the number of sites of the 1D spatial lattice, to go from the non- DTQW-based digitization to the DTQW-based one (or vice versa) -- while this would be linearly costly if the mapping was of near-neighbors type.
In App.\ \ref{app:Gauge}, we show how to include a U(1) gauge-field coupling in our scheme, Eq.\ (6) of the paper, which gives back the known models of Hamiltonian LGT and continuum field theory when taking, respectively, the continuous-time limit and an additional continuous-space limit.
This U(1)-gauged scheme is gauge invariant on the spacetime lattice, and the gauge transformations involve, notably, the \emph{standard} finite differences used in LGT, instead of the more complicated ones used in Ref.\ \cite{AD16}.
We suggest a gauge-invariant quantity on the spacetime lattice and a classical on-shell dynamics (Maxwell's equations) for it, more appropriate than that of Ref.\ \cite{AD16}.}
%\end{center}
\end{widetext} 
%\par
}

\noindent
{\bfseries Note before reading} \\

The reader may have noticed that in the main text, we have essentially reserved the notion of `ultralocality' to evolution operators. 
We did so in order to avoid confusions. 
However, this notion can be used in an abstract, mathematical sense, for an arbitrary operator.
Indeed, given an operator acting on wavefunctions defined on a 1D spatial lattice (e.g., a Hamiltonian or an evolution operator), that is, in practice, a large matrix in the position basis, we qualify it as `ultralocal' if its matrix elements strictly vanish above a certain distance $D_{\text{loc.}}$ from the diagonal\footnote{We use the word `ultralocal' rather than `local' in order to indicate that we exclude `interactions' (the physical content of this word depends on the type of operator) that, e.g., decrease exponentially with the distance, which are also referred to as local in certain contexts \cite{Patel2005}. We stress that this is intended to mean, not that the results presented here cannot be extended to non-ultralocal interactions, but merely that we have not considered this situation. Note the following: here, we work with an infinite number of spatial-lattice sites; in practice, this number is finite, and can vary depending on our amount of resources; by definition of the notion of ultralocality, the ultralocality distance $D_{\text{loc.}}$ does not grow with the number of sites.}.  Notice that the terminology `ultralocal' is equivalent to `near-neighbors'.
Notice also that the physical meaning of the ultralocality depends a priori on the type of operator which is qualified as ultralocal.

Before reading App.\ \ref{app:Wilson_term}, we recommend reading at least Apps.\ \ref{subapp:presentation} and   \ref{subapp:First_comp}.

\section{Unitary and ultralocal digitization of naive fermions with DTQWs} \label{app:naive_case}

This digitization is simply based on the following decomposition of the naive transport term into the sum,
\begin{equation}
H_{\text{n}}^{\text{t}} =  \frac{1}{2} \left[ (H^{\text{t}})' + H^{\text{t}}  \right] \, ,
\end{equation}
of the left-right transport term, $H^{\text{t}}$, already introduced in the main text, and of a right-left one,
\begin{equation} \label{eq:right-left}
(H^{\text{t}})' = \sum_p (H^{\text{t}}_p)' \, ,
\end{equation}
with
{\renewcommand*{\arraystretch}{1}
\begin{align}
 \label{eq:transport_term_bis}
(H_p^{\text{t}})' = \frac{-i }{a} \begin{bmatrix}
0 & \ket{p} \! \! \bra{p+1} -  \ket{p} \! \! \bra{p} \\
\ket p \! \! \bra{p} -  \ket{p+1} \! \! \bra{p} & 0
\end{bmatrix} \, .
\end{align}}
We have straightforwardly that
\begin{equation}
(H^{\text{t}}_p)' = (H^{\text{t}}_p)^{\lozenge} \equiv \sigma^1  H^{\text{t}}_p \,  \sigma^1 \, .
\end{equation}
The operation ${\lozenge}$ thus corresponds to the exchange of the up and down components of the Dirac wavefunction, which is by the way the action of $\gamma^5$ here.
Now, it is easy to check (i) that $(H^{\text{on}}_p)^{\lozenge}$ is, as $H^{\text{on}}_p$, block-diagonal in the non-staggered position basis, so that so is its exponential, but (ii) that $(H^{\text{int.}}_p)^{\lozenge}$ is, on the contrary to  $H^{\text{int.}}_p$, \emph{not} block-diagonal in the staggered position basis. However, it is straightforward to compute the exponential of some operator $O^{\lozenge}$ if we know the exponential of $O$, since using the power-series representation of the exponential together with the fact that $(\sigma^1)^2=1$ immediately gives
\begin{equation}
e^{O^{\lozenge}} = \sigma^1 e^{O} \sigma^1 \equiv (e^{O})^{\lozenge} \, ,
\end{equation}
i.e.\ the operation ${\lozenge}$ commutes with the exponentiation. Applying this to $O = - i \Delta t  H^{\text{int.}}$ first shows, without needing to explicitate the computation, that, because the evolution operator asociated to $H^{\text{int.}}$ is ultralocal, then so is that associated to $(H^{\text{int.}})^{\lozenge}$, which is in the end enough for our purpose. We can thus define an appropriate ultralocal evolution operator associated to $(H^{\text{t}})^{\lozenge}/2$, by (pay attention to the order)
\begin{equation}
(\mathbf{U}^{\text{t}}_{2a})' \equiv (\mathbf{U}^{\text{int.}}_{2a})' (\mathbf{U}^{\text{on}}_{2a})'  \, , 
\end{equation}
where, for $\text{i} = \text{on}, \text{int.}$, we have defined
\begin{equation}
(\mathbf{U}^{\text{i}}_{2a})' \equiv e^{-i \Delta t (\mathbf{H}^{\text{i}})^{\lozenge}/2} =  \left( e^{-i \Delta t \mathbf{H}^{\text{i}}/2}  \right)^{\lozenge}  \equiv (\mathbf{U}^{\text{i}}_{2a})^{\lozenge} \, ,
\end{equation}
where the $\mathbf{U}^{\text{i}}_{a}$'s are exactly the operators that have been introduced in the main text but without indicating the subscript $a$. We immediately obtain
\begin{equation}
(\mathbf{U}^{\text{t}}_{2a})' = (\mathbf{U}^{\text{t}}_{2a})^{\lozenge} \, , 
\end{equation}
where, again, $\mathbf{U}^{\text{t}}_{a}$ is exactly the operator that has been introduced in the main text but without indicating the subscript $a$. Applying the operation $\lozenge$ to Eq.\ (4) considered for $2a$ instead of $a$ results in 
{\small
\renewcommand*{\arraystretch}{1}
\begin{align} \label{eq:Utbiss}
(\mathbf{U}^{\text{t}}_{2a})' =  \left[ 
\begin{array}{*{9}{@{}C{1cm}@{}}}
\tilde{c}^2 & \tilde{s}\tilde{c} & \tilde{s}^2 & -\tilde{s}\tilde{c} &\cdot& \cdot \\
 - \tilde{s}\tilde{c} & \tilde{c}^2 & \cdot & \cdot &\cdot& \cdot  \\
\cdot & \cdot & \tilde{c}^2 &  \tilde{s}\tilde{c} & \tilde{s}^2 &  -\tilde{s}\tilde{c} \\
\tilde{s}\tilde{c} &  \tilde{s}^2  & -\tilde{s}\tilde{c}& \tilde{c}^2 &  \cdot & \cdot   \\
\cdot &\cdot& \cdot &  \cdot & \tilde{c}^2 &  \tilde{s}\tilde{c} \\
 \cdot &\cdot& \tilde{s}\tilde{c}  & \tilde{s}^2 & - \tilde{s}\tilde{c} & \tilde{c}^2
\end{array}
\right] \, ,
\end{align}}
where
\begin{equation}
\tilde{c} = \cos \tilde{\delta}\,  , \ \ \ \  \tilde{s} = \sin \tilde{\delta} \, ,
\end{equation}
with
\begin{equation} \label{eq:delta_tilde}
\tilde{\delta} =\frac{\delta}{2} \, .
\end{equation}
Now, this $(\mathbf{U}^{\text{t}}_{2a})'$ can be written as the following DTQW,
\begin{equation}
(\mathbf{U}^{\text{t}}_{2a})' = S^R_{\mathbf{k}} C(-\tilde{\theta})  S^L_{\mathbf{k}} C(\tilde{\theta}) \, ,
\end{equation}
where
\begin{equation} \label{eq:theta_tilde}
\tilde{\theta} = \pi - 2 \tilde{\delta} \, .
\end{equation}
Finally, we define an appropriate ultralocal evolution operator for the naive transport Hamiltonian, by (pay attention to the order)
\begin{equation}
\mathbf{U}^{\text{t}}_{\text{n}} \equiv (\mathbf{U}^{\text{t}}_{2a})' \mathbf{U}^{\text{t}}_{2a} \, ,
\end{equation}
that is,
\begin{equation}
\mathbf{U}^{\text{t}}_{\text{n}} = \Big[ S^R_{\mathbf{k}} C(-\tilde{\theta})  S^L_{\mathbf{k}} C(\tilde{\theta})   \Big] \Big[ 
C(-\tilde{\theta})  S^R_{\mathbf{k}} C(\tilde{\theta}) S^L_{\mathbf{k}} \Big] \, ,
\end{equation}
which simplifies into
\begin{equation} \label{eq:final_naive}
\mathbf{U}^{\text{t}}_{\text{n}} = S^R_{\mathbf{k}} C(-\tilde{\theta})  S_{\mathbf{k}} C(\tilde{\theta}) S_{\mathbf{k}} (S^R_{\mathbf{k}})^{-1}\, ,
\end{equation}
where
{\renewcommand*{\arraystretch}{1}
\begin{equation} \label{eq:shift_standard}
S_{\mathbf{k}} = S^L_{\mathbf{k}} S^R_{\mathbf{k}} = \begin{bmatrix}
e^{i \mathbf{k}}  & 0 \\ 0 & e^{-i \mathbf{k}} 
\end{bmatrix} \, .
\end{equation}}
Of course, we can also implement a mass term by defining
\begin{equation} \label{eq:non-chiral_dig}
\mathbf{U}_{\text{n}}  \equiv \mathbf{U}^{\text{m}}_{\text{n}}   \mathbf{U}^{\text{t}}_{\text{n}}  \, ,
\end{equation}
where
\begin{equation} \label{eq:mass_term}
\mathbf{U}^{\text{m}}_{\text{n}} \equiv \exp\left[ -i \Delta t \sum_p m \alpha_{\text{n}}^{0}    |p\rangle \! \langle p| \right] \, ,
\end{equation}
and $\alpha_{\text{n}}^{0}$ is any two by two matrix compatible, in the sense of the Clifford algebra, with the choice $\alpha^1 = \sigma^1$ made for the naive transport term, for example any Pauli matrix but $\sigma^1$.

In App.\ \ref{app:chiral}, we briefly comment on how the terms of the continuous-time naive LGT Dirac dynamics, \emph{which exhibit no trace of lattice-chiral transport}\footnote{\emph{Chiral transport} is a type of transport which has some chirality feature, i.e.\ which somehow makes the internal components of the wavefunction move in opposite directions, typically, by choosing the appropriate internal- Hilbert-space basis. We call \emph{lattice-chiral transport} a type of transport which has some chirality feature \emph{in addition to that already coming from the continuous-space limit}, that is, a chirality feature \emph{in the way one performs the spatial discretization}.}, are implemented by the DTQW-based digitization described in the present appendix.

\section{DTQW-based digitization of the continuous-time Wilson term} \label{app:Wilson_term}

The Wilson term,  introduced in the main text, can be viewed as the sum of two terms: a first (resp.\ second) term which, in position space, is diagonal (resp.\ of nearest-neighbors type, and so not `even' block-diagonal),
\begin{subequations}
\begin{align}
(H_{\text{d.}}^{(r)})_p &= \alpha^0 \frac{r}{2a}  \big(   2 \ket{p} \! \! \bra{p} \big)  \, \\
(H_{\text{n.n.}}^{(r)})_p &= - \alpha^0 \frac{r}{2a}  \big(    \ket{p} \! \! \bra{p+1} + \ket{p+1} \! \! \bra{p}  \big)  \, , \label{eq:EQ2}
\end{align}
\end{subequations}
where `d.' (resp.\ `n.n.') stands for `diagonal' (resp.\ `nearest neighbors'). The diagonal term is, as before, irrelevant in the following, since its exponential is ultralocal.

We want to build an ultralocal evolution operator for the nearest-neighbors term. Can we proceed for this term, $H_{\text{n.n.}}^{(r)}$, as we did in App.\ \ref{app:naive_case} for the naively-discretized free Dirac Hamiltonian, $H_{\text{n}}^{\text{t}}$, i.e. build the desired ultralocal evolution operator via a splitting of $H_{\text{n.n.}}^{(r)}$ into the sum of similar left-right and right-left terms? No, and this because of two different reasons: first, $H_{\text{n}}^{\text{t}}$ is a \emph{transport term}, which is traced by the minus sign in $ \ket{p} \! \! \bra{p+1} - \ket{p+1} \! \! \bra{p} $, while $H_{\text{n.n.}}^{(r)}$ is, in the continuum, a \emph{mass term}, which is traced by the plus sign in $\ket{p} \! \! \bra{p+1} + \ket{p+1} \! \! \bra{p}$; second, even if we had a transport term, with a minus sign, instead of a `mass term', with a plus sign, the choice $\alpha^0 = \sigma^3$ in $H_{\text{n.n.}}^{(r)}$ is not suitable for the abovementioned splitting, i.e.\ what would be $\alpha^1$ (and not $\alpha^0 $) if we had a transport term instead of a `mass term', should be antidiagonal, e.g.\ $\sigma^1$ or $\sigma^2$.

That being said, $H_{\text{n.n.}}^{(r)}$ is still ultralocal, and with an ultralocality radius of 1 lattice spacing, so one can still digitize it via a standard even-odd splitting, such as that performed for $H_{\text{n}}^{\text{t}}$ in App.\ \ref{app:non-chiral}. Now, if, instead of performing this even-odd digitization directly on $H_{\text{n.n.}}^{(r)}$, we first express the latter with the Pauli matrix $\sigma^1$, which appears in $H_{\text{n}}^{\text{t}}$, thanks to the relation
\begin{equation} \label{eq:GG}
\sigma^3 = G \sigma^1 G^{-1} \, ,
\end{equation}
where
{\renewcommand*{\arraystretch}{1}
\begin{equation}
G = e^{i \sigma^2 \pi /4} = \frac{1}{\sqrt{2}} \begin{bmatrix}
1 & 1 \\ -1 & 1
\end{bmatrix} \, ,
\end{equation}}
we naturally end up with an even-odd digitization of $H_{\text{n.n.}}^{(r)}$, that we denote by $\mathbf{U}^{(r)}_{\text{e.o.}}$, which, up to a unitarity equivalence induced by Eq.\ (\ref{eq:GG}), (i) is similar to that, $\mathbf{U}^{\text{t}}_{\text{e.o.}}$, of $H_{\text{n}}^{\text{t}}$, and (ii) enables to infer a  DTQW-based digitization, denoted by $\mathbf{U}^{(r)}_{\text{w}}$ (the `w' is for `walk'), via the same transformation as that relating $\mathbf{U}^{\text{t}}_{\text{e.o.}}$ to $\mathbf{U}^{\text{t}}_{\text{n}}$, see Eq.\ (\ref{eq:THE_matrix}).

The computations are equivalent to replacing a (position-space) $\sigma^2$ appearing in $H_{\text{n}}^{\text{t}}$ by a $-r\sigma^1$, which appears in $H_{\text{n.n.}}^{(r)}$\footnote{Indeed, replace $-i\big(\ket{p} \! \! \bra{p+1} - \ket{p+1} \! \! \bra{p}\big)$ by $-r\big(\ket{p} \! \! \bra{p+1} + \ket{p+1} \! \! \bra{p}\big)$.}, and $\mathbf{U}^{(r)}_{\text{w}}$ is thus naturally \emph{defined} by the following counterpart to Eq.\  (\ref{eq:THE_matrix}),
{\begin{widetext}
{\footnotesize
\renewcommand*{\arraystretch}{1}
\begin{equation} \label{eq:THE_matrix_Wilson}
\mathbf{U}^{(r)}_{\text{e.o.}} = \ \ G \left[ 
\begin{array}{*{12}{@{}C{0.8cm}@{}}}
{c}_r^2 & \cdot &\cdot & i{s}_r{c}_r &  - {s}_r^2 & \cdot & \cdot &\cdot & \cdot & \cdot & \cdot & \cdot  \\
\cdot & {c}_r^2 & i{s}_r{c}_r & \cdot &  \cdot &{\color{blue} \boxed{{\color{black}- {s}_r^2}}}  & \cdot &\cdot & \cdot & \cdot &\cdot& \cdot \\
\cdot & i{s}_r{c}_r & {c}_r^2 & \cdot &\cdot & i{s}_r{c}_r & {\color{blue}-s^2} & \cdot & \cdot & \cdot & \cdot & \cdot \\
i{s}_r{c}_r &  \cdot & \cdot & {c}_r^2 & i{s}_r{c}_r & \cdot & \cdot &  \cdot  & \cdot & \cdot & \cdot & \cdot \\
\cdot & \cdot & \cdot &i{s}_r{c}_r &  {c}_r^2 & \cdot & \cdot & i{s}_r{c}_r & -{s}_r^2 &\cdot & \cdot & \cdot \\
\cdot & {\color{blue}-{s}_r^2} & i{s}_r{c}_r & \cdot & \cdot & {c}_r^2 & i{s}_r{c}_r & \cdot & \cdot & {\color{blue}\boxed{{\color{black}-{s}_r^2}}} & \cdot & \cdot \\ 
\cdot &\cdot& {\color{blue}\boxed{{\color{black}-{s}_r^2}}} & \cdot & \cdot & i{s}_r{c}_r &{c}_r^2 & \cdot & \cdot & i{s}_r{c}_r & \color{blue} - {s}_r^2 & \cdot \\
\cdot & \cdot & \cdot & -{s}_r^2 &i{s}_r{c}_r &  \cdot & \cdot & {c}_r^2 &i{s}_r{c}_r &\cdot & \cdot &  \cdot  \\ 
\cdot &  \cdot  & \cdot &  \cdot &  \cdot  & \cdot & \cdot &i{s}_r{c}_r &  {c}_r^2 & \cdot & \cdot &i{s}_r{c}_r \\
\cdot & \cdot & \cdot & \cdot & \cdot & {\color{blue}-{s}_r^2} & i{s}_r{c}_r &  \cdot& \cdot & {c}_r^2 & i{s}_r{c}_r & \cdot \\
\cdot &\cdot & \cdot & \cdot & \cdot & \cdot & {\color{blue} \boxed{{\color{black}-{s}_r^2}}}  & \cdot & \cdot & i{s}_r{c}_r & {c}_r^2 & \cdot \\
\cdot & \cdot & \cdot & \cdot & \cdot & \cdot & \cdot & -{s}_r^2 & i{s}_r{c}_r &  \cdot& \cdot & {c}_r^2 
\end{array}
\right] G^{-1}  \ \ {\color{blue} =  \mathbf{U}^{(r)}_{\text{w}}}  \, ,
\end{equation}}
\end{widetext}}
\noindent
with
\begin{equation}
c_r = \cos{\tilde{\delta}_r} \, , \ \ \ \ s_r = \sin{\tilde{\delta}_r}  \, ,
\end{equation}
where
\begin{equation} \label{eq:deltaR}
\tilde{\delta}_r = r \tilde{\delta} \, ,
\end{equation}
and $\tilde{\delta}$ is given by Eq.\ (\ref{eq:delta_tilde}).
 
Now, as announced,  $G^{-1} \mathbf{U}^{(r)}_{\text{w}} G$ can be written as DTQW similar to $\mathbf{U}^{\text{t}}_{\text{n}}$ in Eq.\ (\ref{eq:final_naive}), namely,
\begin{equation} \label{eq:final_Wilson}
G^{-1} \mathbf{U}^{(r)}_{\text{w}} G = S^R_{\mathbf{k}} K({\tilde{\theta}_r})  S_{\mathbf{k}} K({\tilde{\theta}_r}) S_{\mathbf{k}} (S^R_{\mathbf{k}})^{-1}\, ,
\end{equation}
where $K(\theta)$ is given by Eq.\ (\ref{eq:K}), and
\begin{equation}
\tilde{{\theta}}_r = \pi - 2 \tilde{\delta}_r \, .
\end{equation}
One can prove that $\mathbf{U}^{(r)}_{\text{w}}$  and $\mathbf{U}^{(r)}_{\text{e.o.}}$ are unitarily equivalent, going through the same steps as those followed to show that $\mathbf{U}^{\text{t}}_{\text{n}}$  and $\mathbf{U}^{\text{t}}_{\text{e.o.}}$ are unitarily equivalent, in App.\ \ref{subapp:Mapping}.

\section{Link with Strauch's connection between DTQWs and CTQWs} \label{app:Strauch}

All the $\theta$'s appearing in this appendix stand for $\tilde{\theta}$'s, where $\tilde{\theta}$ is given by Eq.\ (\ref{eq:theta_tilde}); we omit the tilde to make the reading more pleasant.

The DTQW operator appearing in Strauch's work \cite{Strauch06b}, that we shall call Strauch's operator, and with which he derived a connection between DTQWs and CTQWs, is the following,
\begin{equation}
O_{\text{Strauch}} \equiv \breve{S}_{\mathbf{k}} K(\theta) \breve{S}_{\mathbf{k}} K(\theta) \, ,
\end{equation}
with
\begin{equation} \label{eq:K}
K(\theta) = i \breve{C}(\theta) \, ,
\end{equation}
and
\begin{subequations}
\begin{align}
\breve{S}_{\mathbf{k}} &=  S_{\mathbf{k}}^{-1} \\
\breve{C}(\theta) &=  e^{-i\sigma^1 \theta/2} \, ,
\end{align}
\end{subequations}
where $S_{\mathbf{k}}$ is given by Eq.\ (\ref{eq:shift_standard}). Note that we can write
\begin{equation}
O_{\text{Strauch}} = - \breve{S}_{\mathbf{k}} \breve{C}(\theta) \breve{S}_{\mathbf{k}} \breve{C}(\theta) \, .
\end{equation}
This operator is nothing but an elementary DTQW operator, $\breve{S}_{\mathbf{k}} K(\theta)$, applied twice.

Now, in order to make a connection with the DTQW operators appearing in the present work, and, more precisely, with $\mathbf{U}^{\text{t}}_{\text{n}}$, Eq.\ (\ref{eq:final_naive}), one must notice the following unitarity equivalence,
\begin{subequations}
\begin{align}
\breve{S}_{\mathbf{k}} &=  P S_{\mathbf{k}} P^{-1} \\
\breve{C}(\theta) &= P {C}(\theta) P^{-1} \, ,
\end{align}
\end{subequations}
where $C(\theta)=e^{-i\sigma^2 \theta/2}$,
and the passage matrix is
{\renewcommand*{\arraystretch}{1}
\begin{align}
P = i \begin{bmatrix}
0 & e^{-i \pi/4} \\ e^{i \pi/4} & 0
\end{bmatrix} =
\begin{bmatrix}
0 & e^{i \pi/4} \\ -e^{-i \pi/4} & 0
\end{bmatrix}  \, .
\end{align}}
With this, and using for $\mathbf{U}^{\text{t}}_{\text{n}}$, Eq.\ (\ref{eq:final_naive}),  the notation
\begin{equation}
\mathbf{U}^{\text{t}}_{\text{n}} = \mathbf{U}^{\text{t}}_{\text{n}}(\theta,\theta) \, ,
\end{equation}
one can easily write
\begin{equation}
O_{\text{Strauch}} = - P S^L_{\mathbf{k}} \, \, \mathbf{U}^{\text{t}}_{\text{n}}(-\theta,\theta)  \, \, (P S^L_{\mathbf{k}})^{-1} \, ,
\end{equation}
This shows that $O_{\text{Strauch}}$ and $-\mathbf{U}^{\text{t}}_{\text{n}}(-\theta,\theta)$ are unitarily equivalent.
Now, that $O_{\text{Strauch}}$ has a continuous-time limit implies that in the limit of a vanishing time step $\Delta t = 0$, it equals the identity, which is manifestly \emph{not} unitarily equivalent to $-O_{\text{Strauch}}$. Hence, this holds for an arbitrary time step $\Delta t $, i.e. $-O_{\text{Strauch}}$ is \emph{not} unitarily equivalent to $O_{\text{Strauch}}$, and is thus not unitarily equivalent to $\mathbf{U}^{\text{t}}_{\text{n}}(-\theta,\theta)$, since the latter \emph{is} unitarily equivalent to $-O_{\text{Strauch}}$.

Taking the limit  $\theta \rightarrow \pi$ makes Strauch's scheme coincide with a standard CTQW on the line, instead of the previously well-known limit $\theta \rightarrow 0$, associated to a continuum limit in both time \emph{and space}, which delivers the Dirac equation with $\sigma^3$ as the first alpha matrix.
In the present work, we use $\mathbf{U}^{\text{t}}_{\text{n}}(\theta,\theta)$ instead of $\mathbf{U}^{\text{t}}_{\text{n}}(-\theta,\theta)$: this is necessary to be able to derive a continuous-space limit after the continuous-time one, which, as with $\theta \rightarrow 0$, makes the scheme coincide with the Dirac equation, but in a different Clifford-algebra representation, with $\sigma^1$ as the first alpha matrix. 

Note that Strauch's proof for the connection between DTQWs and CTQWs \emph{already} introduces the fact that one applies \emph{twice} the shift and coin operations: this appears to be a necessary ingredient to derive a continuous-time limit from a standard DTQW in Strauch's (and hence in the present) framework.
Indeed, Strauch's procedure to take the continuous-time limit can be viewed as a (time-)continuum limit in the fashion of Ref.\ \cite{DMD13b}, i.e.\ a (time-)continuum limit performed with a stroboscope of period 2. 
This comes from the following technical analogy between Strauch's time-continuum limit and the spacetime-continum limit of Ref.\ \cite{DMD13b}: in Ref.\ \cite{DMD13b}, increasing the stroboscope period from 1 to 2 enables to release the constraint on the zeroth-order value of the mixing angle, i.e.\ no need for it to be homogeneous in spacetime anymore, for the continuum limit to exist; in Strauch's (and hence in the present) work, this procedure enables to release the constraint on the spatial smoothness of the lattice wavefunction, i.e.\ no need to take a continuous-space limit as we take a continuous-time limit.

\section{A non- DTQW-based digitization of continuous-time naive fermions}
\label{app:non-chiral}

All the $\theta$'s appearing in this appendix stand for $\tilde{\theta}$'s, where $\tilde{\theta}$ is given by Eq.\ (\ref{eq:theta_tilde}); we omit the tilde to make the reading more pleasant, and this also enables us to use the tilde for another purpose, namely, to denote the Fourier transform.

\subsection{Presentation}
\label{subapp:presentation}

The non- DTQW-based digitization is the standard even-odd straightforward (one could say `naive') one: the naive transport lattice Hamiltonian is split into even and odd parts,
\begin{equation}
H_{\text{n}}^{\text{t}} = H_{\text{e.o.}}^{\text{e}} + H_{\text{e.o.}}^{\text{o}} \, ,
\end{equation}
where `e.o.' is for `even-odd', that is, it indicates that the digitization is non- DTQW-based, and where
\begin{subequations}
\begin{align}
H^{\text{e}}_{\text{e.o}} &= \sum_l (H^{\text{t}}_{\text{n}})_{2l} \\ 
H^{\text{o}}_{\text{e.o.}} &=  \sum_l (H^{\text{t}}_{\text{n}})_{2l+1} \, ,
\end{align}
\end{subequations}
which are block-diagonal in the position basis, so that so are their exponentials, which is the trick used to preserve ultralocality in the time discretization. 

The Trotter formula enables to write the evolution from time $j$ to time $j+1$ as (we now work with matrix representations, but use, for state vectors, the same braket notation)
\begin{align}
|\Phi_{j+1}\rangle = \mathbf{U}_{\text{e.o.}} |\Phi_j\rangle  + O(\Delta t) \, ,
\end{align}
where, instead of the DTQW-based digitization, Eq.\ (\ref{eq:non-chiral_dig}), we here consider
\begin{equation} \label{eq:chiral_dig}
\mathbf{U}_{\text{e.o.}} = \mathbf{U}^{\text{m}}_{\text{n}} \mathbf{U}^{\text{t}}_{\text{e.o.}} \, ,
\end{equation}
with
\begin{equation}
\mathbf{U}^{\text{t}}_{\text{e.o.}} =  \mathbf{U}^{\text{e}}_{\text{e.o.}} \mathbf{U}^{\text{o}}_{\text{e.o.}}\,
\end{equation}
and
\begin{equation}
\mathbf{U}^{\text{e}/\text{o}}_{\text{e.o.}} = e^{-i \Delta t \mathbf{H}^{\text{e}/\text{o}}_{\text{e.o.}}} \, ,
\end{equation}
and where the mass term is still given by Eq.\ (\ref{eq:mass_term}).
After a computation analog to that performed in the paper, we end up with
{\small
\renewcommand*{\arraystretch}{1}
\begin{subequations}
\begin{align}
\mathbf{U}^{\text{e}}_{\text{e.o.}} &= \left[
\begin{array}{*{9}{@{}C{1cm}@{}}}
c & -\sigma^1s & \cdot &\cdot& \cdot & \cdot \\
\sigma^1s & c & \cdot & \cdot & \cdot & \cdot \\
\cdot & \cdot & c & -\sigma^1s & \cdot & \cdot  \\
\cdot & \cdot & \sigma^1s & c & \cdot & \cdot \\
\cdot &\cdot& \cdot & \cdot & c & -\sigma^1s \\
\cdot &\cdot& \cdot & \cdot &  \sigma^1s & c 
\end{array}
\right] \\
\mathbf{U}^{\text{o}}_{\text{e.o.}} &=   \left[ 
\begin{array}{*{9}{@{}C{1cm}@{}}}
c & \cdot & \cdot & \cdot &\cdot& \cdot \\
\cdot & c & -\sigma^1s & \cdot &\cdot& \cdot  \\
\cdot &  \sigma^1s& c & \cdot &\cdot & \cdot  \\
\cdot & \cdot &\cdot &c & -\sigma^1s & \cdot  \\
\cdot &  \cdot &\cdot &\sigma^1s& c & \cdot  \\
\cdot & \cdot & \cdot & \cdot &\cdot & c
\end{array}
\right] \, ,
\end{align}
\end{subequations}}
where, for a good visibility, we have written $6\times6$ matrices (of $2\times 2$ matrices) in the (non-staggered) position basis, $(|p\rangle)_{p\in\mathbb{Z}}$, i.e.\ $12 \times 12$ matrices in the staggered position basis, $(|n\rangle)_{n\in\mathbb{Z}}$, that is, as big as that of Eq.\ (\ref{eq:THE_matrix_Wilson}). The product reads
{\small
\renewcommand*{\arraystretch}{1}
\begin{align} \label{eq:Utbis-naive}
\mathbf{U}^{\text{t}}_{\text{e.o.}} =   \left[ 
\begin{array}{*{9}{@{}C{1cm}@{}}}
c ^2 & -\sigma^1 sc & s^2 & \cdot &\cdot& \cdot \\
\sigma^1 sc & c^2 & -\sigma^1sc & \cdot &\cdot& \cdot  \\
\cdot &  \sigma^1sc& c^2 &  -\sigma^1 sc & s^2 &  \cdot \\
\cdot &  s^2 & \sigma^1sc& c^2 &  -\sigma^1 sc & \cdot    \\
\cdot &\cdot& \cdot &  \sigma^1 sc & c ^2 & -\sigma^1 sc \\
 \cdot &\cdot& \cdot  & s^2 & \sigma^1sc & c^2
\end{array}
\right] \, ,
\end{align}}
\noindent
which is translationally invariant \emph{only every two sites} of the (non-staggered) position basis.

\iffalse
For the sake of completeness, let us explicitly write the action of $\mathbf{U}_{\text{e.o.}}$:
{\small
{\begin{subequations}\label{eq:U}
\begin{align}
\Phi_{j+1,p} &= \left( \mathbf{U}_{\text{e.o.}} \Phi_{j} \right)_p \\
& \hspace{-0.4cm} =  M \left( c^2  \Phi_{j,p} - \sigma^1 sc \, ( \Phi_{j,p+1} -   \Phi_{j,p-1} ) + s^2  \Phi_{j,p+2(-1)^p} \right) \, ,
\end{align}
\end{subequations}}
\normalsize
\noindent
with
\begin{equation}
M = \begin{bmatrix}
\mu & 0 \\ 0 & \mu^{\ast}
\end{bmatrix} . 
\end{equation}
\fi

\subsection{First comparisons with the DTQW-based digitization} \label{subapp:First_comp}

Let us compare the non- DTQW-based digitization, $\mathbf{U}^{\text{t}}_{\text{e.o.}}$, given by Eq.\ (\ref{eq:Utbis-naive}), to the DTQW-based, $\mathbf{U}^{\text{t}}_{\text{n}}$ (we recall that the `n' is for `naive'), given by Eq.\ (\ref{eq:final_naive}).

\subsubsection{First comment, on the compact writings}
From now on, we will often consider the infinite-dimensional matrices as linear operators acting on the wavefunction $\Psi_j : p \mapsto \Psi_{j,p}$, rather than on the infinite column vector $|\Psi_j)$, and will use, instead of the notation $|\Phi_{j+1}\rangle = \mathbf{U}_{\text{e.o.}} |\Phi_j\rangle  + O(\Delta t) $, the writing $\Psi_{j+1,p} = \left( \mathbf{U} \Psi_{j} \right)_p + O(\Delta t) $, without changing the notation $\mathbf{U}$ used for the infinite-dimensional matrix -- the context should make it clear whether the distinction matters or not. 

The one- time-step evolution equations induced by $\mathbf{U}_{\text{e.o.}}$ and $\mathbf{U}_{\text{n}}$ respectively read:
{\small
{\begin{subequations}\label{eq:UW}
\begin{align}
\Phi_{j+1,p} &= \left( \mathbf{U}_{\text{e.o.}} \Phi_{j} \right)_p \label{eq:U} \\
& \hspace{-0.4cm} =  M \left[ c^2  \Phi_{j,p} - \sigma^1 sc \, ( \Phi_{j,p+1} -   \Phi_{j,p-1} ) + s^2  \Phi_{j,p+2(-1)^p} \right] \, ,  \nonumber \\
\Psi_{j+1,p} &= \left( \mathbf{U}_{\text{n}} \Psi_{j} \right)_p \label{eq:Wbis} \\
& \hspace{-0.4cm} =  M \left[ c^2  \Psi_{j,p} - \sigma^1 sc \, ( \Psi_{j,p+1} -   \Psi_{j,p-1} ) + s^2  \big( S_{\mathbf{k}}^2 \Psi_{j} \big)_p \right] \, , \nonumber 
\end{align}
\end{subequations}}
\normalsize
\noindent
with
{\renewcommand*{\arraystretch}{1}
\begin{equation}
M = \begin{bmatrix}
\mu & 0 \\ 0 & \mu^{\ast}
\end{bmatrix} \, ,
\end{equation}}
and where we recall that $S_{\mathbf{k}}$ is an internal-state-dependent shift, given by Eq.\ (\ref{eq:shift_standard}).

Now, we have
\begin{align}
c &= 1 + O(\Delta t^2) \nonumber \\ 
s &= O(\Delta t) \, ,
\end{align}
so that the $s^2$ term of Eqs. (\ref{eq:U}) and (\ref{eq:Wbis}) vanishes in the continuous-time limit. One could multiply $s^2$ by any factor as long as the evolution remains unitary and ultralocal, which are the main requirements in the present work; Eqs (\ref{eq:U}) and (\ref{eq:Wbis}) are two such possibilities. One may say that, in order for the ultralocal scheme to be unitary,  (i) the non- DTQW-based digitization uses as a factor of $s^2$ a position-dependent shift, and more precisely a site-parity- dependent shift, while (ii) the DTQW-based digitization uses an internal-state dependent shift. In App.\ \ref{app:EO_DTQW}, we show that the even-odd scheme can actually be seen as a staggered version of a scheme which \emph{is} DTQW-based, \emph{not} with the $LR$ coin basis, but with an even-odd coin basis that one can introduce.

\subsubsection{A more detailed comparison}

We have, in the staggered position basis,
{\begin{widetext}
{\small
\renewcommand*{\arraystretch}{1}
\begin{equation} \label{eq:THE_matrix}
\mathbf{U}^{\text{t}}_{\text{e.o.}} = \left[ 
\begin{array}{*{12}{@{}C{0.7cm}@{}}}
c^2 & \cdot &\cdot & -sc &  s^2 & \cdot & \cdot &\cdot & \cdot & \cdot & \cdot & \cdot  \\
\cdot & c^2 & -sc & \cdot &  \cdot &{\color{blue} \boxed{{\color{black}s^2}}}  & \cdot &\cdot & \cdot & \cdot &\cdot& \cdot \\
\cdot & sc & c^2 & \cdot &\cdot & -sc & {\color{blue}s^2} & \cdot & \cdot & \cdot & \cdot & \cdot \\
sc &  \cdot & \cdot & c^2 & -sc & \cdot & \cdot &  \cdot  & \cdot & \cdot & \cdot & \cdot \\
\cdot & \cdot & \cdot & sc &  c^2 & \cdot & \cdot & -sc & s^2 &\cdot & \cdot & \cdot \\
\cdot & {\color{blue}s^2} & sc & \cdot & \cdot & c ^2 & -sc & \cdot & \cdot & {\color{blue}\boxed{{\color{black}s^2}}} & \cdot & \cdot \\ 
\cdot &\cdot& {\color{blue}\boxed{{\color{black}s^2}}} & \cdot & \cdot & sc & c^2 & \cdot & \cdot & -sc & \color{blue} s^2 & \cdot \\
\cdot & \cdot & \cdot & s^2 & sc &  \cdot & \cdot & c ^2 & -sc &\cdot & \cdot &  \cdot  \\ 
\cdot &  \cdot  & \cdot &  \cdot &  \cdot  & \cdot & \cdot & sc &  c^2 & \cdot & \cdot & -sc \\
\cdot & \cdot & \cdot & \cdot & \cdot & {\color{blue}s^2} & sc &  \cdot& \cdot & c ^2 & -sc & \cdot \\
\cdot &\cdot & \cdot & \cdot & \cdot & \cdot & {\color{blue} \boxed{{\color{black}s^2}}}  & \cdot & \cdot & sc & c^2 & \cdot \\
\cdot & \cdot & \cdot & \cdot & \cdot & \cdot & \cdot & s^2 & sc &  \cdot& \cdot & c ^2 
\end{array}
\right] {\color{blue} =  \mathbf{U}^{\text{t}}_{\text{n}} }  \, ,
\end{equation}}
\end{widetext}}
\noindent
with the following color code: all the matrix elements of  $\mathbf{U}^{\text{t}}_{\text{e.o.}}$ are in black (the blue $s^2$'s must be replaced by zeros), while those of $\mathbf{U}^{\text{t}}_{\text{n}}$ are those in black, omitting those in blue boxes, which must replaced by zeros, and adding those in blue (which were zeros in  $\mathbf{U}^{\text{t}}_{\text{e.o.}}$). Going from $\mathbf{U}^{\text{t}}_{\text{e.o.}}$ to $\mathbf{U}^{\text{t}}_{\text{n}}$ restores translational invariance (recall that $\mathbf{U}^{\text{t}}_{\text{e.o.}}$ is translationally invariant only every two sites). 

Now, $\mathbf{U}^{\text{t}}_{\text{e.o.}}$ and $\mathbf{U}^{\text{t}}_{\text{n}}$ are two infinite-dimensional unitary matrices. 
Since the usual finite-dimensional- case spectral theorem can be extended to the infinite-dimensional case for unitary operators, both $\mathbf{U}^{\text{t}}_{\text{e.o.}}$ and $\mathbf{U}^{\text{t}}_{\text{n}}$ are diagonalizable. 
Hence, a necessary and sufficient condition for them to be unitarily equivalent, is that they have the same eigenvalues.
The answer, positive, is easily given by a formal computation software: the $12\times 12$ sub-blocks of $\mathbf{U}^{\text{t}}_{\text{e.o.}}$ and $\mathbf{U}^{\text{t}}_{\text{n}}$ given by Eq.\ (\ref{eq:THE_matrix}), have the same eigenvalues; one then concludes invoking the common every-two-sites translational invariance of both infinite-dimensional matrices.
In the next section, we make use of this every-two-site translational invariance to show explicitly, by going to Fourier space, that $\mathbf{U}^{\text{t}}$ and $\mathbf{W}^{\text{t}}$ are unitarily equivalent.

\subsection{Mapping between the DTQW-based and the non- DTQW-based digitizations} \label{subapp:Mapping}

\subsubsection{In Fourier space} \label{subsubapp:in_Fourier_space}

$\mathbf{U}^{\text{t}}_{\text{n}}$ is translationally invariant on the lattice $\{x_p=pa, p \in \mathbb{Z} \}$, but $\mathbf{U}^{\text{t}}_{\text{e.o.}}$ only every two lattice sites, so we compare both evolutions by combining two consecutive lattice sites in the following way.
Consider from now on that $p=2l$.
We define a Fourier transform every two sites,
\begin{equation} \label{eq:Fourier}
\widetilde{\chi}_{(K)} = \sum_{l\in \mathbb{Z}} \chi_l \, e^{-iKl}\, , 
\end{equation}
where we have used the notation $K=2k$, and $k\in[-\pi,\pi[$ is the wavevector. We have, see Eq.\ (\ref{eq:U}) with $m=0$,
{\small
\renewcommand*{\arraystretch}{1.2}
\begin{equation} \label{eq:EO2}
\left( {\mathbf{U}^{\text{t}}}^{(2)}_{\! \! \! \text{e.o.}} \Phi^{(2)} \right)_l = 
\begin{bmatrix}
c^2 \Phi^E_{l} - \sigma^1 sc \left( \Phi^O_{l} - \Phi^O_{l-1} \right) + s^2 \Phi^E_{l+1} \\
c^2 \Phi^O_{l} - \sigma^1 sc \left( \Phi^E_{l+1} - \Phi^E_{l} \right) + s^2 \Phi^O_{l-1} 
\end{bmatrix} ,
\end{equation}}
where ${\mathbf{U}^{\text{t}}}^{(2)}_{\text{e.o.}}$ is the redefinition of $\mathbf{U}^{\text{t}}_{\text{e.o.}}$ (viewed as an operator and not a matrix) as acting on
{\renewcommand*{\arraystretch}{1.1}
\begin{equation}
 \Phi^{(2)}: l \mapsto \Phi^{(2)}_l = \begin{bmatrix}
\Phi_{2l} \\ \Phi_{2l+1} 
\end{bmatrix}
= \begin{bmatrix}
\Phi^{E}_{l} \\ \Phi^{O}_{l} 
\end{bmatrix} ,
\end{equation}}
so that, expanded in the internal-Hilbert-space basis, the Fourier representation $\mathcal{U}^{(2)}$ of ${\mathbf{U}^{\text{t}}}^{(2)}_{\! \! \! \text{e.o.}}$ is a $4\times4$ matrix-multiplication operator, i.e.\ acting as 
\begin{equation}
\left( \mathcal{U}^{(2)} \widetilde{\Phi}^{(2)} \right)_{(K)} = \mathcal{U}^{(2)}_{(K)} \widetilde{\Phi}^{(2)}_{(K)} \, ,
\end{equation}
with components
{\begin{widetext}
\renewcommand*{\arraystretch}{1}
\begin{equation}
\mathcal{U}^{(2)}_{(K)} = 
\begin{bmatrix}
c^2 + s^2 e^{iK} & \cdot & \cdot & -sc (1-e^{-iK}) \\
\cdot & c^2 + s^2 e^{iK} & -sc (1-e^{-iK}) &  \cdot \\
\cdot & -sc (e^{iK}-1) &  c^2 + s^2 e^{-iK}&  \cdot \\
 -sc (e^{iK}-1) & \cdot &  \cdot & c^2 + s^2 e^{-iK}
\end{bmatrix} .
\end{equation}
The corresponding matrix for $\mathbf{U}^{\text{t}}_{\text{n}}$ is (see Eq.\ (5))
\begin{equation} \label{eq:Wmathcal}
\mathcal{W}^{(2)}_{(K)} = 
\begin{bmatrix}
c^2 + s^2 e^{iK} & \cdot & \cdot & -sc (1-e^{-iK}) \\
\cdot & c^2 + s^2 e^{-iK} & -sc (1-e^{-iK}) &  \cdot \\
\cdot & -sc (e^{iK}-1) &  c^2 + s^2 e^{iK}&  \cdot \\
 -sc (e^{iK}-1) & \cdot &  \cdot & c^2 + s^2 e^{-iK}
\end{bmatrix} ,
\end{equation}
\end{widetext}}
\noindent
where the `$\mathcal{W}$' stands for `walk' (we could have used the notation `$\mathbf{W}_{\text{n}}$' instead of `$\mathbf{U}_{\text{n}}$', since it is the DTQW-based digitization).

$\mathcal{U}^{(2)}_{(K)}$ and $\mathcal{W}^{(2)}_{(K)}$ are two $4 \times 4$ unitary matrices. 
They are equal in the $2 \times 2$ subspace formed by the first and last columns and rows. It is thus enough to focus on the complementary subspace, formed by the second and third columns and rows,
{\renewcommand*{\arraystretch}{1}
\begin{subequations}
\begin{align}
\Pi \big(\mathcal{U}^{(2)}_{(K)}\big) &= 
\begin{bmatrix}
 c^2 + s^2 e^{iK} & -sc (1-e^{-iK})  \\
-sc (e^{iK}-1) &  c^2 + s^2 e^{-iK}
\end{bmatrix}  \\
\Pi \big(\mathcal{W}^{(2)}_{(K)}\big) &= 
\begin{bmatrix}
c^2 + s^2 e^{-iK} & -sc (1-e^{-iK})  \\
-sc (e^{iK}-1) &  c^2 + s^2 e^{iK}
\end{bmatrix} .
\end{align}
\end{subequations}}
The above two matrices are unitary and thus diagonalizable (as well as the original $4\times 4$ ones).
It turns out that they have the same eigenvalues (and hence the original $4\times 4$ ones as well).
Hence, $\mathcal{U}^{(2)}_{(K)}$ and $\mathcal{W}^{(2)}_{(K)}$ are unitarily equivalent,  i.e.\ there exist $\mathcal{B}_{(K)}$ unitary such that
\begin{equation}
\mathcal{B}_{(K)} \mathcal{U}^{(2)}_{(K)} \mathcal{B}_{(K)}^{-1} = \mathcal{W}^{(2)}_{(K)} ,
\end{equation}
and $\mathcal{B}_{(K)}$ is explicitly given, for example, by
\begin{equation} \label{eq:BK}
\mathcal{B}_{(K)} = \mathcal{Q}_{(K)} \mathcal{P}_{(K)}^{-1}\, ,
\end{equation}
where $\mathcal{P}_{(K)}$ (resp.\ $\mathcal{Q}_{(K)}$) is the matrix whose (i) first and last columns and rows are those of 
$\mathcal{U}^{(2)}_{(K)}$ (resp.\ $\mathcal{W}^{(2)}_{(K)}$), and (ii) second and third columns and rows are formed by the normalized eigenvectors $\Pi \big( \mathcal{U}^{(2)}_{(K)} \big)$ (resp.\ $\Pi \big( \mathcal{W}^{(2)}_{(K)} \big)$). In particular, $\mathcal{P}_{(K)}$ (resp.\ $\mathcal{Q}_{(K)}$) is thus unitary.

\subsubsection{In real space}

Given our definition of the Fourier transform, Eq.\ (\ref{eq:Fourier}), we have
\begin{equation}
\Psi_l^{(2)} = \frac{1}{2\pi} \int_{-\pi}^{\pi} dk  \tilde{\Psi}_{(K)}^{(2)} e^{iKl} \, ,
\end{equation}
so that
\begin{subequations}
\begin{align}
\Psi_l^{(2)} &= \frac{1}{2\pi} \int_{-\pi}^{\pi} dk  \mathcal{B}_{(K)} \tilde{\Phi}_{(K)}^{(2)} e^{iKl}  \\
&= \frac{1}{2\pi} \int_{-\pi}^{\pi}  dk \mathcal{B}_{(K)} \left( \sum_{l'\in \mathbb{Z}} \Phi_{l'}^{(2)}  e^{-iKl'}\right)  e^{iKl} \, ,
\end{align}
\end{subequations}
that is,
\begin{align}
\Psi_l^{(2)} = \sum_{l'\in \mathbb{Z}} \mathbf{B}_{ll'} \Phi_{l'}^{(2)} \, ,
\end{align}
with
\begin{equation}
\mathbf{B}_{ll'} = \mathbf{b}_{l-l'} = \frac{1}{2\pi} \int_{-\pi}^{\pi}  dk \mathcal{B}_{(K)}  e^{iK(l-l')} \, ,
\end{equation}
which is, as $\mathcal{B}_{(K)}$, a $4 \times 4$ matrix, with coefficients
\begin{equation} \label{eq:coeffs}
\mathbf{B}_{ll'}^{uv} = \mathbf{b}_{l-l'}^{uv} = \frac{1}{2\pi} \int_{-\pi}^{\pi}  dk \mathcal{B}_{(K)}^{uv}  e^{iK(l-l')} \, .
\end{equation}

We wonder whether the mapping $\mathbf{B}$ is ultralocal, i.e., whether its coefficients $\mathbf{B}_{ll'} = \mathbf{b}_{l-l'}$ vanish when  we are sufficiently far from the diagonal, i.e.\ for $|l-l'|$ big enough.
\iffalse
independently, of course, of the size of $\mathbf{B}$. Since here we work with an infinite-dimensional position Hilbert space, no size appears in computations, and so none of the following computations will depend on such a size.
\fi

Since $\mathbf{b}_{l-l'}$ is a $4 \times 4$ matrix, it is convenient to reformulate the question as whether there is at least one of its coefficients  $\mathbf{b}_{l-l'}^{uv}$ which do not vanish for $|l-l'|$ big enough, i.e.\ whether there is at least one of the $\mathbf{b}^{uv} : N \mapsto \mathbf{b}^{uv}_{N}$ of $\mathbb{Z} \rightarrow \mathbb{C}$ which has \emph{not} a finite support on $\mathbb{Z}$. 
Now, inverting Eq.\ (\ref{eq:coeffs}) yields
\begin{equation} \label{eq:genpol}
\mathcal{B}_{(K)}^{uv} = \sum_{N'\in \mathbb{Z}} \mathbf{b}^{uv}_{N'} e^{-iKN'}\, ,
\end{equation}
so that $\mathbf{b}^{uv}$ has a finite support on $\mathbb{Z}$ if and only the above sum involves a finite number of terms.

Because of what we said in the last paragraph of Sec.\ \ref{subsubapp:in_Fourier_space}, we know a priori that $\mathcal{B}_{(K)}$, which is unitary and defined in Eq.\ (\ref{eq:BK}), has the following form,
{\renewcommand*{\arraystretch}{1}
\begin{equation} \label{eq:BKexpanded}
\mathcal{B}_{(K)} = 
\begin{bmatrix}
1 & \cdot & \cdot & \cdot \\
\cdot & \ast & \ast & \cdot \\
\cdot & \ast& \ast &\cdot \\
\cdot & \cdot & \cdot & 1
\end{bmatrix} \, ,
\end{equation}}
where the four asterisks stand for
{\renewcommand*{\arraystretch}{1.2}
\begin{equation} \label{eq:BK22}
\Pi \big( \mathcal{B}_{(K)} \big) = 
\begin{bmatrix}
a_{(K)} & b_{(K)}\\
 - e^{i\varphi} b_{(K)}^{\ast} & e^{i\varphi} a_{(K)}^{\ast}  \\
\end{bmatrix} \, ,
\end{equation}}
which we know can be written has a generic $2\times2$ unitary matrix, and determined pedestrianly, i.e.\ via the product
\begin{equation} \label{eq:BK22product}
\Pi \big( \mathcal{B}_{(K)} \big) = \mathcal{S}_{(K)} \mathcal{R}_{(K)}^{-1} \, ,
\end{equation}
where 
$\mathcal{R}_{(K)}$ (resp.\ $\mathcal{S}_{(K)}$) is the matrix whose columns are formed by the normalized eigenvectors of 
$\Pi \big( \mathcal{U}^{(2)}_{(K)} \big)$ (resp.\ $\Pi \big( \mathcal{W}^{(2)}_{(K)} \big)$). After some simplifications, we find
\begin{subequations}
\begin{align}
\varphi &= 0 \\
a_{(K)} &= 2 F  \\
b_{(K)} &= F \left(  \tan(\tilde{\delta}) ( 1+ e^{-iK} ) \right) \, ,
\end{align} 
\end{subequations}
where
\begin{align}
F = \frac{1}{2} (1 + X)^{-\frac{1}{2}}\label{eq:F} \, ,
\end{align}
and
\begin{align}
X = \frac{\tan^2(\tilde{\delta})}{4} \left( 2 + e^{iK} + e^{-iK} \right)  \, . \label{eq:finalll}
\end{align}

Now, it is manifest from the following writing of  $\mathcal{B}_{(K)}$,
{\small
\renewcommand*{\arraystretch}{1}
\begin{equation} \label{eq:B_final}
\mathcal{B}_{(K)} =  F
\begin{bmatrix}
\frac{1}{F} & \cdot & \cdot & \cdot \\
\cdot & 2 & \tan(\tilde{\delta})  (1+  e^{-iK})  & \cdot \\
\cdot & - \tan(\tilde{\delta}) (1+  e^{iK}) & 2 & \cdot \\
\cdot & \cdot & \cdot & \frac{1}{F}
\end{bmatrix} \, ,
\end{equation}}
\noindent
that the latter will be a generalized (negative powers allowed) polynomial of $e^{iK}$, i.e.\ of the form of Eq.\ (\ref{eq:genpol}) with a sum involving a finite number of terms,  if and only if $F$ is also one.
Since $|X| < 1$, we have, for any $\omega \in \mathbb{R}$,
\begin{equation}
(1+X)^{\omega} = \sum_{n=0}^{+\infty} \left( \frac{1}{n!} \prod_{k=0}^{n-1} (\omega - k) \right) X^n \, .
\end{equation}
Applying this formula for $\omega = -1/2$ shows that $F$ is \emph{not} reducible to a (generalized) polynomial, i.e.\ it is an integer series of $X$ involving an infinite number of terms, and so the mapping between the non- DTQW-based and the DTQW-based discretizations is \emph{not} ultralocal.

\section{Towards $\mathrm{U(1)}$ lattice gauge theory with DTQWs.}
\label{app:Gauge}

\subsection{Left-right scheme}

We now work with the abstract position basis, $(\ket{p})_{p\in\mathbb{Z}}$.
Instead of the evolution operator given by Eqs.\ (6), consider now
\begin{equation} \label{eq:compact_gauged}
\hat{U}^{\text{g;t}}= e^{-i\hat{\alpha}_j} C(-\theta) S^R_{\hat{k}} S^R_{\hat{\vartheta}_j} C(\theta)  S^L_{\hat{\vartheta}_j} S^L_{\hat{k}} \, ,
\end{equation}
where `g' is for `gauged', and $\hat{\alpha}_j$ and $\hat{\vartheta}_j$ are operators which are diagonal in the position basis, that is,
$
\hat{\alpha}_j \ket p =  \alpha_{j,p} \ket p\equiv  \Delta t \, q A^0_{j,p} \ket p,$ and
$\hat{\vartheta}_j \ket p = \vartheta_{j,p} \ket p \equiv - a \, q A^1_{j+1,p} \ket p,
$
with real-valued eigenvalues. 
%
\iffalse
Note that one can write 
$
S^L_{\hat{\vartheta}_j} S^L_{\hat{k}} = S^L_{\hat{k}} T_{\hat{k}}(S^L_{\hat{\vartheta}_j}),
$
where we have introduced the translation operation by one lattice site to the right,
$
T_{\hat{k}}(\hat{O}) = e^{-i\hat{k}} \hat{O}  e^{i\hat{k}}.
$
\fi
In the continuum-spacetime limit, one gets the Dirac equation with an electric 2-potential coupling $(A^0,A^1)$ through the charge $q$, as we are going to show by first taking only the continuous-time limit.
In terms of the equations of motion, replacing Evolution (6) by Evolution (\ref{eq:compact_gauged}) means replacing Eqs.\ (5) (we omit the mass) by 
{\footnotesize
\begin{align} \label{eq:W_gauged}
\psi_{j+1,p}^{L} &= e^{- i \alpha_{p}} \left( sc \, e^{-i \vartheta_{p-1}} \psi_{p-1}^{R} + c^2 \psi^{L}_{p} - sc \, \psi_{p}^{R} + s^2e^{i \vartheta_p} \psi_{p+1}^{L} \right)   \nonumber \\
\psi_{j+1,p}^{R} &=  e^{- i \alpha_{p}} \left(  s^2 e^{- i \vartheta_{p-1}} \psi_{p-1}^{R} + sc  \, \psi_{p}^{L} + c^2 \psi^{R}_{p} - sc \, e^{i \vartheta_p} \psi_{p+1}^{L} \right) \, .
\end{align}}
\noindent
The continuous-time limit of the above equations yields a Hamiltonian evolution with what we shall call the left-right \emph{gauged} Hamiltonian, whose single-site (transport) term is
{\small$
\hat{H}_p^{\text{g;t}} = (-i /a) \ \text{antidiag}( \ket{p} \! \! \bra{p} -  \ket{p+1} \! \! \bra{p} e^{- i \hat{\vartheta}_j},
\ket p \! \! \bra{p+1} e^{i \hat{\vartheta}_j} -  \ket{p} \! \! \bra{p}) + q \hat{A}^0_j \ket{p} \! \! \bra{p}.
$}
Taking now the continuous-space limit, one ends up with the announced Dirac equation.

The gauged evolution, Eqs.\ (\ref{eq:W_gauged}), is invariant under discrete-spacetime local gauge transformations, that is, invariant under $\Psi_{j,p} \rightarrow e^{iq\varphi_{j,p}} \Psi_{j,p}$, where $\varphi_{j,p} \in \mathbb{R}$ is an arbitrary local phase, provided the gauge field transforms as
\begin{subequations} \label{eq:gauge_transfo}
\begin{align}
A^0_{j,p}  &\longrightarrow A^0_{j,p} -\frac{1}{\Delta t} \left( \varphi_{j+1,p} - \varphi_{j,p} \right) \label{eq:A0} \\
A^1_{j,p}  &\longrightarrow  A^1_{j,p} + \frac{1}{a} \left( \varphi_{j,p+1} - \varphi_{j,p} \right) \, . \label{eq:A1}
\end{align}
\end{subequations}
In the continuous-time limit, the transformation on the gauge field reduces to
$
A^0_{p}  \rightarrow A^0_{p} - \frac{d}{dt} \varphi_p, $ and
$A^1_{p}   \rightarrow  A^1_{p} + \frac{1}{a} \left( \varphi_{p+1} - \varphi_{p} \right).
$
Taking the continuous-space limit of the previous transformation, one recovers the standard gauge transformation of an Abelian Yang-Mills gauge field,
$
A_{\mu} \rightarrow  A_{\mu} - \partial_{\mu} \phi.
$
Notice that the above gauge transformation on the spacetime lattice, which is given by standard finite differences, is extremely simple with respect to that of Ref.\ \cite{DMD14}. This ensues from the fact that scheme of the present work, Eq.\  (6), is chirally symmetrized with respect to that of Ref.\ \cite{DMD14}, i.e., viewing this from another perspective, symmetrized in time: indeed, in the present two-time-step scheme, the second step is done with an opposite coin angle $-\theta$, i.e.\ with a coin operator which is the time-symmetrized of that having angle $\theta$.
\iffalse
, which yields, in the continuous-time limit, a Hamiltonian-LGT (or, more generally, a tight-binding-like) standard transport term \emph{which is Hermitean} (both terms $p+1 \rightarrow p$ and $p \rightarrow p+1$ appear).
\fi

The following quantity,
\begin{equation}
(F_{01})_{j,p} \equiv (d_0 A_1)_{j,p} - (d_1 A_0)_{j,p} \, ,
\end{equation}
where, for any quantity $Q_{j,p}$ defined on the spacetime lattice, we have introduced $
(d_0 Q)_{j,p} = (Q_{j+1,p} - Q_{j,p})/\Delta t$ and
$
(d_1 Q)_{j,p} = (Q_{j+1,p+1} - Q_{j+1,p})/a,
$
is invariant under the transformation on the gauge field in Eqs. (\ref{eq:gauge_transfo}) (since $d_0$ and $d_1$ commute) and tends, in the continuous-spacetime limit, towards the standard electric tensor,
$
{F}_{01} = \partial_0 A_1 - \partial_1 A_0.
$
The following alternate quantity,
\begin{equation} \label{eq:transporter}
(U_{01})_{j,p} = e^{iqa^2 (F_{01})_{j,p}} \, ,
\end{equation}
is actually a more appropriate gauge-invariant quantity, because, as the equations of motion, Eqs.\ (\ref{eq:W_gauged}), are, it is invariant under the transformation $A^{\mu}_{j,p} \rightarrow A^{\mu}_{j,p} + 2w^{\mu}_{j,p} \pi / (q \Delta^{\mu})$, with $\Delta^{0}=\Delta t$ and $\Delta^{1}=a$, and $w^{\mu}_{j,p}\in\mathbb{Z}$ such that $w^1_{j+1,p} - w^1_{j,p} \neq - (w^0_{j+1,p+1} - w^0_{j+1,p})$, whereas $({F}_{01})_{j,p}$ is not.
This quantity, Eq.\ (\ref{eq:transporter}), is exactly that considered in Euclidean (or Wilson), and hence discrete- (imaginary) time (Abelian) LGTs, see Eq.\ (5.20) of Ref.\ \cite{book_Rothe}.
Note that one can simply use $U_{01}$ instead of  $F_{01}$ for possible discrete equivalents to Maxwell's equations in the fashion of Ref.\ \cite{AD16}, since this does not modify the continuum limit, given that what appears in those equations are the (discrete) \emph{derivatives} of  $U_{01}$.

\subsection{Naive scheme}

We have seen that Eq.\ (\ref{eq:compact_gauged}) is a possible U(1)-gauged version of Eq.\ (6). Now, a possible U(1)-gauged version of Eq.\ (\ref{eq:final_naive}), is
\begin{align}
U_{\text{n}}^{\text{g;t}} = e^{-i\hat{\alpha}_j} S^R_{\hat{k}} C^{\text{g}}(-\tilde{\theta}, \hat{\vartheta}_j)  S_{\hat{k}} C^{\text{g}}(\tilde{\theta}, \hat{\vartheta}_j) S_{\hat{k}} (S^R_{\hat{k}})^{-1}\, ,
\end{align}
where
{\renewcommand*{\arraystretch}{1}
\begin{equation}
C^{\text{g}}(\tilde{\theta}, \hat{\vartheta}_j)  =
\begin{bmatrix}
e^{i\hat{\vartheta}_j}  \cos \frac{\tilde{\theta}}{2} & - \sin  \frac{\tilde{\theta}}{2} \\
\sin \frac{\tilde{\theta}}{2}  & e^{-i\hat{\vartheta}_j}  \cos \frac{\tilde{\theta}}{2} 
\end{bmatrix} \, .
\end{equation}}

The continuous-time limit of the evolution induced by $U_{\text{n}}^{\text{g;t}}$ is a Hamiltonian dynamics given by the gauged version of the massless naive Hamiltonian, that is,
\begin{align} \label{eq:naive_transport_gauged}
(H_{\text{n}}^{\text{g;t}})_p &= \frac{-i}{2a} \alpha^1 \big(  \ket{p} \! \! \bra{p+1}  e^{i \hat{\vartheta}_{p}} -  \ket{p+1} \! \! \bra{p}  e^{-i \hat{\vartheta}_{p}}\big)  \nonumber  \\ 
& \ \ \ + q \hat{A}_p^{0} \ket p \! \! \bra{p}  \, .  
\end{align}

\section{Comment on the DTQW-based digitization of the continuous-time \emph{naive} LGT Dirac dynamics} \label{app:chiral}

Let us comment on the particularity of the digitization of the continuous-time \emph{naive} LGT Dirac dynamics through DTQW.
Consider the following walk operator,
\begin{equation} \label{eq:compact_DTQW_2_n}
\mathsf{W}^{\text{t}}(-\tilde{\theta}_1,\tilde{\theta}_2) = S^R_{\mathbf{k}} C(-\tilde{\theta}_2) S_{\mathbf{k}} C(\tilde{\theta}_1)  S_{\mathbf{k}} (S^R_{\mathbf{k}})^{-1} \, ,
\end{equation}
which is nothing but $\mathbf{U}^{\text{t}}_{\text{n}}$, given by Eq.\ (\ref{eq:final_naive}), but considering it as a function of two variables
\begin{equation}
\tilde{\theta}_i = \pi - 2 \tilde{\delta}_i\, ,
\end{equation}
$i=1,2$, with
\begin{equation}
\tilde{\delta}_i = \kappa_i \frac{\Delta t}{2a} \, .
\end{equation}
The parameter $\kappa_i$ is introduced to keep a trace, in the calculations, of where the various terms come from in the original discrete-time scheme. 
Expanding the above compact writing, Eq.\ (\ref{eq:compact_DTQW_2_n}), yields the following one- time-step evolution equations,
{\small
\begin{subequations}
\begin{align} \label{eq:WW}
\psi_{j+1,p}^{L} &=  \tilde{s}_2 \tilde{s}_1 \psi_{p+2}^{L} + \tilde{c}_2 \tilde{c}_1 \psi^{L}_{p} - \tilde{c}_2\tilde{s}_1 \, \psi^{R}_{p+1} + \tilde{s}_2\tilde{c}_1 \, \psi^{R}_{p-1}    \\
\psi^{R}_{ j+1,p} &= \tilde{c}_2 \tilde{c}_1 \psi^{R}_{p} - \tilde{s}_2\tilde{c}_1 \, \psi_{p+1}^{L} + \tilde{c}_2\tilde{s}_1 \, \psi_{p-1}^{L} +  \tilde{s}_2\tilde{s}_1 \psi^{R}_{p-2} , \! \!
\end{align}
\end{subequations}}
which correspond essentially to Eq.\ (\ref{eq:Wbis}), but with
{\small
\begin{subequations}
\begin{align}
\tilde{s}_i &= \cos \frac{\tilde{\theta}_i}{2} = \sin \tilde{\delta}_i = \tilde{\delta}_i + O(\tilde{\delta}_i^3) =  \kappa_i \frac{\Delta t}{2 a} + O(\Delta t^3)\\
\tilde{c}_i &= \sin \frac{\tilde{\theta}_i}{2} = \cos \tilde{\delta}_i = 1 + O(\tilde{\delta}_i^2) = 1 + O(\Delta t^2)  \, .
\end{align}
\end{subequations}}
The continuous-time limit of this scheme reads
{\renewcommand*{\arraystretch}{1.2}
\begin{equation} \label{eq:rep_mix_2}
\dot{\Psi}_p = - \frac{1}{2a}
\begin{bmatrix}
\kappa_1 \psi^{R}_{p+1} - \kappa_2 \psi^{R}_{p-1} \\
 \kappa_2 \psi^L_{p+1} - \kappa_1 \psi^L_{p-1} 
\end{bmatrix} \, .
\end{equation}}

As announced, the $\kappa_i$'s enable to visualize how the terms of the continuous-time dynamics are implemented by the discrete-time automaton, and  make it manifest, in the continuous-time limit, the lattice-chiral aspect of the discrete-time implementation.

\section{Even-odd digitization as a DTQW-based digitization by introducing an even-odd coin space.} \label{app:EO_DTQW}

The DTQW writing of Eqs.\ (5), namely, Eq.\ (6), can actually be understood in the staggered picture, by replacing \(L\) (resp.\ \(R\)) by `even' (resp.\ `odd') and by using the staggered-picture lattice, but this demands to be able to realize, on this lattice, (two-site) translations of even-site (resp.\ odd-site) components without translating the odd-site (resp. even-site) ones. In other words, if such translations can be realized, no single-site translations are needed to evolve the walker on the staggered-picture lattice, and the suggested procedure is conceptually equivalent to the left-right picture since it naturally introduces an even-odd ($EO$) internal Hilbert space. Such an even-odd picture with even-odd internal degree of freedom should thus simply be seen as a possible instance of the left-right picture, with possible experimental interest. 

To sum up the previous paragraph: the staggered picture of the left-right- Hamiltonian digitization can be viewed as a possible instance of the non-staggered picture, \emph{provided one can realize site-parity- dependent two-site translations on the staggered-picture lattice}. We are going to show that, similarly, the even-odd digitization (referred to as non- DTQW-based) of naive fermions, presented in App.\ \ref{app:non-chiral}, can also be viewed, under the same condition, as such a DTQW in the $EO$ coin basis (with, of course, an \emph{additional} $LR$ internal degree of freedom on which \emph{no} DTQW is performed).

Eq.\ (\ref{eq:EO2}) can be rewritten as
{\renewcommand*{\arraystretch}{1.2}
\begin{align} \label{eq:EO2QW}
&\left( {\mathbf{U}^{\text{t}}}^{(2)}_{\! \! \! \text{e.o.}} \Phi^{(2)} \right)_l =  \\ 
&\ \ \ \ \left[ \begin{array}{rr}
\tilde{s} \ \ \left( \tilde{s}  \Phi^E_{l+1} - \tilde{c} \sigma^1  \Phi^O_{l} \right) & + \ \tilde{c} \sigma^1 \left(  \tilde{c} \sigma^1 \Phi^E_{l} +  \tilde{s} \Phi^O_{l-1} \right)  \\
- \tilde{c} \sigma^1 \left( \tilde{s} \Phi^E_{l+1}  - \tilde{c} \sigma^1 \Phi^O_{l} \right) &+ \ \ \tilde{s} \ \ \left(  \tilde{c}  \sigma^1 \Phi^E_{l} + \tilde{s} \Phi^O_{l-1} \right) 
\end{array} \right] \, , \nonumber 
\end{align}}
that is to say, 
{\renewcommand*{\arraystretch}{1}
\begin{align} \label{eq:EOQW}
{U^{\text{t}}}^{(2)}_{\! \! \! \text{e.o.}} = 
\begin{bmatrix}
\tilde{s} & -\tilde{c} \sigma^1 \\ \tilde{c} \sigma^1 & \tilde{s}
\end{bmatrix} 
\begin{bmatrix}
1 & 0 \\ 0  & e^{-i\hat{K}}
\end{bmatrix}
\begin{bmatrix}
\tilde{s} & \tilde{c} \sigma^1 \\ - \tilde{c} \sigma^1 & \tilde{s}
\end{bmatrix}
\begin{bmatrix}
e^{i\hat{K}} & 0 \\ 0  & 1
\end{bmatrix} \, ,
\end{align}}
with $\hat{K} = 2 \hat{k}$, and where the (block-)matrices are written in the $EO$ coin basis, that we are allowed to introduce provided one can realize site-parity- dependent two-site translations, which makes single-site translations unnecessary (and forbidden if we introduce the $EO$ coin basis). The scheme may be rewritten as
\begin{align} \label{eq:EOQW2}
{U^{\text{t}}}^{(2)}_{\! \! \! \text{e.o.}} =  \mathsf{V} \,  \mathsf{C}(-\tilde{\theta})
\mathsf{S}^{R}_{\hat{K}}
\mathsf{C}(\tilde{\theta})
\mathsf{S}^{L}_{\hat{K}}
 \, \mathsf{V}^{\dag} \, ,
\end{align}
with $\mathsf{C}(\theta)$ and  $\mathsf{S}^{L/R}_{\hat{K}}$ matricially given by $C(\theta)$ and  $S^{L/R}_{\hat{K}}$, respectively,  but where the {\textbackslash}mathsf font indicates that the coin basis is here the $EO$ one, not the $LR$ one. We have also introduced the following site-parity- dependent change of $LR$ coin basis,
{\renewcommand*{\arraystretch}{1}
\begin{equation}
\mathsf{V} = \begin{bmatrix}
\rho & 0 \\ 0 &  \rho^{\dag} 
\end{bmatrix} \, ,
\end{equation}}
where 
{\renewcommand*{\arraystretch}{1}
\begin{equation}
\rho = e^{i\frac{\pi}{4}}\begin{bmatrix}
1 & -i \\ -i &  1
\end{bmatrix} \, ,
\end{equation}}
is a square root of the Pauli matrix $\sigma^1$, i.e. $\rho^2 = \sigma^1$.

\end{document}